\documentclass[10pt]{article}
\usepackage{graphicx}
\usepackage{amsmath}
\usepackage{amssymb}
\usepackage{caption2}
\begin{document}
\bibliographystyle{prsty}
\begin{center}
{\large {\bf \sc{  Analysis of  the hidden-charm  pentaquark molecular states with strangeness and without strangeness  via the QCD sum rules }}} \\[2mm]
Zhi-Gang Wang \footnote{E-mail: zgwang@aliyun.com.  }, Qi Xin      \\
 Department of Physics, North China Electric Power University, Baoding 071003, P. R. China
\end{center}

\begin{abstract}
In this article, we investigate the $\bar{D}\Sigma_c$, $\bar{D}\Xi^\prime_c$,    $\bar{D}\Sigma_c^*$, $\bar{D}\Xi_c^*$, $\bar{D}^{*}\Sigma_c$, $\bar{D}^{*}\Xi^\prime_c$,
  $\bar{D}^{*}\Sigma_c^*$ and $\bar{D}^{*}\Xi_c^*$ pentaquark molecular states with strangeness and without strangeness via  the QCD sum rules at length,  pay much attention to the light flavor $SU(3)$ breaking effects,  and make predictions for new pentaquark molecular states besides assigning the $P_c(4312)$, $P_c(4380)$, $P_c(4440)$, $P_c(4457)$ and $P_{cs}(4459)$ self-consistently. We can search for those pentaquark molecular states in the decays of the $\Lambda_b^0$, $\Xi_b^0$ and $\Xi_b^-$ in the future. Furthermore, we discuss the higher dimensional vacuum condensates in details.
\end{abstract}



\section{Introduction}

In 2015,  the  LHCb collaboration explored the $\Lambda_b^0\to J/\psi K^- p$ decays and observed  two pentaquark candidates $P_c(4380)$ and $P_c(4450)$ in the $J/\psi p$  mass spectrum with the  preferred quantum numbers  $J^P={\frac{3}{2}}^-$ and  ${\frac{5}{2}}^+$, respectively \cite{LHCb-4380}.
The  Breit-Wigner   masses and widths are  $M_{P_c(4380)}=4380\pm 8\pm 29\,\rm{MeV}$, $M_{P_c(4450)}=4449.8\pm 1.7\pm 2.5\,\rm{MeV}$, $\Gamma_{P_c(4380)}=205\pm 18\pm 86\,\rm{MeV}$, and  $\Gamma_{P_c(4450)}=39\pm 5\pm 19\,\rm{MeV}$, respectively.
In 2019,  the LHCb collaboration re-investigated  a data sample of the $\Lambda_b^0\to J/\psi K^- p$ decays, which was an
order of magnitude larger than that previously analyzed, and  observed a  narrow pentaquark candidate $P_c(4312)$ in the $ J/\psi  p$ mass spectrum, and confirmed the structure $P_c(4450)$, which are consisted  of two narrow overlapping peaks $P_c(4440)$ and $P_c(4457)$ \cite{LHCb-Pc4312}.
   The measured  Breit-Wigner masses and widths are
\begin{flalign}
 &P_c(4312) : M = 4311.9\pm0.7^{+6.8}_{-0.6} \mbox{ MeV}\, , \, \Gamma = 9.8\pm2.7^{+ 3.7}_{- 4.5} \mbox{ MeV} \, , \nonumber \\
 & P_c(4440) : M = 4440.3\pm1.3^{+4.1}_{-4.7} \mbox{ MeV}\, , \, \Gamma = 20.6\pm4.9_{-10.1}^{+ 8.7} \mbox{ MeV} \, , \nonumber \\
 &P_c(4457) : M = 4457.3\pm0.6^{+4.1}_{-1.7} \mbox{ MeV} \, ,\, \Gamma = 6.4\pm2.0_{- 1.9}^{+ 5.7} \mbox{ MeV} \,   .
\end{flalign}
 Very recently, the LHCb collaboration reported an evidence of a hidden-charm pentaquark candidate $P_{cs}(4459)$ with  strangeness $S=-1$ in the $J/\psi \Lambda$ invariant mass spectrum with  a statistical significance of  $3.1\sigma$ in the $\Xi_b^- \to J/\psi K^- \Lambda$ decays   using the $pp$ collision data corresponding to a total integrated luminosity of
$9\rm{ fb}^{-1}$ collected with the LHCb experiment at centre-of-mass energies of 7, 8 and 13 TeV \cite{LHCb-Pcs4459-2012},
the Breit-Wigner mass and width are
\begin{flalign}
 &P_{cs}(4459) : M = 4458.8 \pm 2.9 {}^{+4.7}_{-1.1} \mbox{ MeV}\, , \, \Gamma = 17.3 \pm 6.5 {}^{+8.0}_{-5.7} \mbox{ MeV} \, ,
\end{flalign}
but the spin and parity have not been determined yet. The  $P_c(4312)$, $P_c(4380)$, $P_c(4440)$, $P_c(4457)$ and $P_{cs}(4459)$ lie slightly below  the thresholds of the  meson-baryon pairs $\bar{D}\Sigma_c$, $\bar{D}\Sigma_c^*$, $\bar{D}^*\Sigma_c$ and $\bar{D}^*\Xi_c$, respectively, the nearby meson-baryon thresholds are shown clearly in Table \ref{Thresholds}, because  the $D$ and $D^*$ mesons, the ${1\over 2}^+$ flavor antitriplet ($\Lambda_c^+(2286)$,  $\Xi_c^+(2468)$, $\Xi_c^0(2471)$),  and the ${1\over 2}^+$ and ${3\over 2}^+$ flavor sextets
($\Omega_c(2695)$,   $\Xi^\prime_c(2579)$,  $\Sigma_c(2453)$) and ($\Omega_c^*(2766)$,  $\Xi_c^*(2646)$,  $\Sigma_c^*(2518)$)
have been well established \cite{PDG}.

As expected, the  $P_c(4312)$, $P_c(4380)$, $P_c(4440)$, $P_c(4457)$ and $P_{cs}(4459)$ have been tentatively assigned to be the $\bar{D}\Sigma_c$, $\bar{D}\Sigma_c^*$, $\bar{D}^*\Sigma_c$, $\bar{D}^*\Sigma_c^*$ and $\bar{D}^*\Xi_c$ pentaquark molecular states respectively based on the contact-range effective field theory \cite{Penta-mole-CREFT,Pcs4459-mole-CREFT},
one-boson exchange potential model \cite{Penta-mole-OBE-CC,Pcs4459-mole-OBE-CC,Pcs4459-mole-OBE-2-CC}, (quasipotential) Bethe-Salpeter equation \cite{Penta-mole-BSE-CC,Pcs4459-mole-BSE-HJ-CC,Pcs4459-mole-BSE-CC,Pcs4459-mole-BSE-2-CC}, effective Lagrangian approach \cite{Penta-mole-ELA,Pcs4459-mole-ELA}, effective-range expansion and resonance compositeness relation \cite{Penta-mole-ERE-CC}, Lippmann-Schwinger equation \cite{Penta-mole-LSE-CC,Penta-mole-LSE-2-CC}, the QCD sum rules \cite{Penta-mole-QCDSR-Chen,Penta-mole-QCDSR-Zhang,Penta-mole-QCDSR-Azizi,WZG-IJMPA-mole,Pcs4459-mole-QCDSR}, etc.

At first glance, $M_{\bar{D}^{*0}}+M_{\Xi_c^0}-M_{P_{cs}(4459)}=19\,\rm{MeV}$, $M_{\bar{D}^0}+M_{\Sigma_c^+}-M_{P_{c}(4312)}=6\,\rm{MeV}$, it is odd that the $\bar{D}^{*0}\Xi_c^0$ molecular state which involving exchanges of the strange mesons is more tightly bound than the $\bar{D}^0\Sigma_c^+$ molecular state which involving exchanges of the non-strange mesons, we have to introduce the coupled-channel effects \cite{Penta-mole-OBE-CC,Pcs4459-mole-OBE-CC,Pcs4459-mole-OBE-2-CC,Penta-mole-BSE-CC,Pcs4459-mole-BSE-HJ-CC,Pcs4459-mole-BSE-CC,Pcs4459-mole-BSE-2-CC,Penta-mole-ERE-CC,Penta-mole-LSE-CC,Penta-mole-LSE-2-CC}.

In the QCD sum rules, we usually choose the local currents to interpolate the tetraquark or pentaquark  molecular states  having two color neutral clusters \cite{Penta-mole-QCDSR-Chen,Penta-mole-QCDSR-Zhang,Penta-mole-QCDSR-Azizi,WZG-IJMPA-mole,Pcs4459-mole-QCDSR}, which are not necessary to be the physical mesons and baryons, and the tetraquark or pentaquark  molecular states are not necessary to be loosely bound, they can be compact objects and can lie below or above the corresponding meson-meson or meson-baryon pairs \cite{WZG-color-neutral}.

On the other hand, the $P_c(4312)$, $P_c(4380)$, $P_c(4440)$, $P_c(4457)$ and $P_{cs}(4459)$ can also be  tentatively assigned to be the diquark-diquark-antiquark type (or diquark-triquark type) pentaquark states in the diquark-model through exploring their masses \cite{di-di-anti-penta-mass-1, di-di-anti-penta-mass-2} (\cite{di-tri-penta-mass}) and decay modes \cite{di-di-anti-penta-decay-1, di-di-anti-penta-decay-2, di-di-anti-penta-decay-3} (\cite{di-tri-penta-width}) via the effective  Hamiltonian, or investigating their masses \cite{Wang1508-EPJC,WangHuang1508-1,WangHuang1508-2,WangHuang1508-3,WZG-Pcs4459}, decays \cite{Pcs4459-penta-di-di-decay}, electromagnetic properties \cite{Pcs4459-penta-di-di-EM}  via the QCD sum rules.

 In Ref.\cite{WZG-WHJ-XQ}, we suggest the  hadronic dressing  mechanism  to compromise the pentaquark and pentaquark molecule interpretations based on the calculations of the QCD sum rules,  the pentaquark states maybe have a diquark-diquark-antiquark type pentaquark core with the typical size of the $qqq$-type  baryon states, the strong couplings  to the meson-baryon pairs lead to some pentaquark molecule Fock components, and the pentaquark states maybe spend  a rather  large time as the  molecular states. We can choose either the diquark-diquark-antiquark type currents or color-singlet-color-singlet type five-quark currents to interpolate the pentaquark states.

\begin{table}
\begin{center}
\begin{tabular}{|c|c|c|c|c|c|c|c|c|c|c|c|}\hline\hline
($\bar{D}^0\,\Xi_c^0$, $\bar{D}^-\,\Xi_c^+$)                       & (4336, 4337)        \\

($\bar{D}^{*0}\,\Xi_c^0$, $\bar{D}^{*-}\,\Xi_c^+$)                 & (4478, 4478)        \\

($\bar{D}^0\,\Sigma_c^+$, $\bar{D}^{-}\,\Sigma_c^{++}$)            & (4318, 4323)        \\
($\bar{D}^{0}\,\Sigma_c^{*+}$, $\bar{D}^{-}\,\Sigma_c^{*++}$)      & (4382, 4388)        \\
($\bar{D}^{*0}\,\Sigma_c^+$, $\bar{D}^{*-}\,\Sigma_c^{++}$)        & (4460, 4464)        \\
($\bar{D}^{*0}\,\Sigma_c^{*+}$, $\bar{D}^{*-}\,\Sigma_c^{*++}$)    & (4524, 4529)        \\

($\bar{D}^0\,\Xi_c^{\prime0}$, $\bar{D}^{-}\,\Xi_c^{\prime+}$)     & (4444, 4448)        \\
($\bar{D}^{0}\,\Xi_c^{*0}$, $\bar{D}^{-}\,\Xi_c^{*+}$)             & (4511, 4515)        \\
($\bar{D}^{*0}\,\Xi_c^{\prime0}$, $\bar{D}^{*-}\,\Xi_c^{\prime+}$) & (4586, 4589)        \\
($\bar{D}^{*0}\,\Xi_c^{*0}$, $\bar{D}^{*-}\,\Xi_c^{*+}$)           & (4653, 4656)        \\  \hline\hline

\end{tabular}
\end{center}
\caption{ The  thresholds of the open-charm meson and baryon pairs, where the unit is MeV.} \label{Thresholds}
\end{table}

In the scenario of the pentaquark molecular states, the $P_c(4312)$, $P_c(4380)$, $P_c(4440)$, $P_c(4457)$ are assigned to the $\bar{D}\Sigma_c$, $\bar{D}\Sigma_c^*$, $\bar{D}^*\Sigma_c$, $\bar{D}^*\Sigma_c^*$ pentaquark molecular states, which involve charmed baryon states in flavor sextets ${\bf 6}_f$ \cite{Penta-mole-CREFT,Penta-mole-OBE-CC,Penta-mole-BSE-CC,
Penta-mole-ELA,Penta-mole-ERE-CC,Penta-mole-LSE-CC,Penta-mole-LSE-2-CC,
Penta-mole-QCDSR-Chen,Penta-mole-QCDSR-Zhang,Penta-mole-QCDSR-Azizi,WZG-IJMPA-mole}, while the $P_{cs}(4459)$ is assigned to be the $\bar{D^*}\Xi_c$ pentaquark molecular state, which involves charmed baryon state in flavor antitriplet ${\bf \bar{3}}_f$ \cite{Pcs4459-mole-CREFT,Pcs4459-mole-OBE-CC,Pcs4459-mole-OBE-2-CC,
Pcs4459-mole-BSE-HJ-CC,Pcs4459-mole-BSE-CC,Pcs4459-mole-BSE-2-CC,Pcs4459-mole-ELA,Pcs4459-mole-QCDSR}. Thus the $P_c(4312/4380/4440/4457)$ and $P_{cs}(4459)$ belong to different flavor multiplets, in the present work, we will focus on the flavor sextets ${\bf 6}_f$.

  In this article, we extend our previous work \cite{WZG-IJMPA-mole} to study the masses and pole residues of the  $\bar{D}\Sigma_c$, $\bar{D}\Xi^\prime_c$,
   $\bar{D}\Sigma_c^*$, $\bar{D}\Xi_c^*$, $\bar{D}^{*}\Sigma_c$, $\bar{D}^{*}\Xi^\prime_c$,
  $\bar{D}^{*}\Sigma_c^*$ and $\bar{D}^{*}\Xi_c^*$ pentaquark molecular states    with the QCD sum rules by accomplishing  the operator product expansion up to the vacuum condensates  of dimension $13$ in a consistent way, and take  account of the vacuum  condensates  $\langle \bar{q}q\rangle\langle \frac{\alpha_s}{\pi}GG\rangle$,
$\langle \bar{q}q\rangle^2\langle \frac{\alpha_s}{\pi}GG\rangle$ and $\langle \bar{q}q\rangle^3\langle \frac{\alpha_s}{\pi}GG\rangle$ neglected in Ref.\cite{WZG-IJMPA-mole},  and revisit the assignments of the $P_c(4312)$, $P_c(4380)$, $P_c(4440)$, $P_c(4457)$ and $P_{cs}(4459)$. Furthermore, we pay much attention to the light flavor $SU(3)$ breaking effects.

 The article is arranged in the form:  we acquire the QCD sum rules for the masses and pole residues of the  pentaquark molecular states  in Sect.2;  in Sect.3, we present the numerical results and discussions; and Sect.4 is reserved for our
conclusion.

\section{QCD sum rules for  the  pentaquark molecular states}

We write down  the two-point correlation functions $\Pi(p)$, $\Pi_{\mu\nu}(p)$ and $\Pi_{\mu\nu\alpha\beta}(p)$  in the QCD sum rules,
\begin{eqnarray}\label{CF-Pi}
\Pi(p)&=&i\int d^4x e^{ip \cdot x} \langle0|T\left\{J(x)\bar{J}(0)\right\}|0\rangle \, , \nonumber\\
\Pi_{\mu\nu}(p)&=&i\int d^4x e^{ip \cdot x} \langle0|T\left\{J_{\mu}(x)\bar{J}_{\nu}(0)\right\}|0\rangle \, , \nonumber\\
\Pi_{\mu\nu\alpha\beta}(p)&=&i\int d^4x e^{ip \cdot x} \langle0|T\left\{J_{\mu\nu}(x)\bar{J}_{\alpha\beta}(0)\right\}|0\rangle \, ,
\end{eqnarray}
where the currents $J(x)=J^{\bar{D}\Sigma_c}(x)$, $J^{\bar{D}\Xi^\prime_c}(x)$, $J_\mu(x)=J^{\bar{D}\Sigma_c^*}_{\mu}(x)$, $J^{\bar{D}\Xi_c^*}_{\mu}(x)$, $ J^{\bar{D}^*\Sigma_c}_{\mu}(x)$, $ J^{\bar{D}^*\Xi^\prime_c}_{\mu}(x)$, $J_{\mu\nu}(x)=J^{\bar{D}^*\Sigma_c^*}_{\mu\nu}(x)$, $J^{\bar{D}^*\Xi_c^*}_{\mu\nu}(x)$,
\begin{eqnarray}\label{current-J}
 J^{\bar{D}\Sigma_c}(x)&=& \bar{c}(x)i\gamma_5 u(x)\, \varepsilon^{ijk}  u^T_i(x) C\gamma_\alpha d_j(x)\, \gamma^\alpha\gamma_5 c_{k}(x) \, ,\nonumber \\
  J^{\bar{D}\Xi^\prime_c}(x)&=& \bar{c}(x)i\gamma_5 u(x)\, \varepsilon^{ijk}  u^T_i(x) C\gamma_\alpha s_j(x)\, \gamma^\alpha\gamma_5 c_{k}(x) \, ,
 \end{eqnarray}

 \begin{eqnarray}
 J^{\bar{D}\Sigma_c^*}_{\mu}(x)&=& \bar{c}(x)i\gamma_5 u(x)\, \varepsilon^{ijk}  u^T_i(x) C\gamma_\mu d_j(x)\, c_{k}(x) \, ,\nonumber \\
 J^{\bar{D}\Xi_c^*}_{\mu}(x)&=& \bar{c}(x)i\gamma_5 u(x)\, \varepsilon^{ijk}  u^T_i(x) C\gamma_\mu s_j(x)\, c_{k}(x) \, ,
  \end{eqnarray}

 \begin{eqnarray}
 J^{\bar{D}^*\Sigma_c}_{\mu}(x)&=& \bar{c}(x)\gamma_\mu u(x)\, \varepsilon^{ijk}  u^T_i(x) C\gamma_\alpha d_j(x)\, \gamma^\alpha\gamma_5 c_{k}(x) \, ,\nonumber \\
    J^{\bar{D}^*\Xi^\prime_c}_{\mu}(x)&=& \bar{c}(x)\gamma_\mu u(x)\, \varepsilon^{ijk}  u^T_i(x) C\gamma_\alpha s_j(x)\, \gamma^\alpha\gamma_5 c_{k}(x) \, ,
  \end{eqnarray}

 \begin{eqnarray}\label{current-Jmunu}
 J^{\bar{D}^*\Sigma_c^*}_{\mu\nu}(x)&=& \bar{c}(x)\gamma_\mu u(x)\, \varepsilon^{ijk}  u^T_i(x) C\gamma_\nu d_j(x)\,   c_{k}(x) +(\mu\leftrightarrow\nu)\, ,\nonumber \\
 J^{\bar{D}^*\Xi_c^*}_{\mu\nu}(x)&=& \bar{c}(x)\gamma_\mu u(x)\, \varepsilon^{ijk}  u^T_i(x) C\gamma_\nu s_j(x)\,   c_{k}(x) +(\mu\leftrightarrow\nu)\, ,
\end{eqnarray}
the $i$, $j$, $k$ are color indices. In the present work, we choose the color-singlet-color-singlet type  currents  $J(x)$,
 $J_{\mu}(x)$ and $J_{\mu\nu}(x)$ to interpolate the  pentaquark molecular states with the spin-parity $J^P={\frac{1}{2}}^-$,  ${\frac{3}{2}}^-$ and ${\frac{5}{2}}^-$, respectively. The currents couple potentially to the pentaquark molecular states having two color neutral clusters,  one has the same quantum numbers as the charmed mesons, the other has the same quantum numbers as the charmed baryons, they are not physical mesons and baryons, as we choose the local five-quark currents, while the mesons and baryons are spatial extended objects and have mean spatial sizes $\sqrt{\langle r^2\rangle} \neq 0$, for example, $\sqrt{\langle r^2\rangle_{E,\Sigma_{c}^{++}}}=0.48\,\rm{fm}$,
$\sqrt{\langle r^2\rangle_{M,\Sigma_{c}^{++}}}=0.83\,\rm{fm}$, $\sqrt{\langle r^2\rangle_{M,\Sigma_{c}^{0}}}=0.81\,\rm{fm}$ from the lattice QCD,
where the subscripts $E$ and $M$ stand for the Electric  and magnetic radii, respectively \cite{Can-Sigmac-2015},
$\sqrt{\langle r^2\rangle_{M,\Sigma_{c}^{++}}}=0.77\,\rm{fm}$, $\sqrt{\langle r^2\rangle_{M,\Sigma_{c}^{0}}}=0.52\,\rm{fm}$, $\sqrt{\langle r^2\rangle_{M,\Sigma_{c}^{+}}}=0.81\,\rm{fm}$, $\sqrt{\langle r^2\rangle_{M,\Xi_{c}^{\prime+}}}=0.55\,\rm{fm}$, $\sqrt{\langle r^2\rangle_{M,\Xi_{c}^{\prime0}}}=0.79\,\rm{fm}$ from the self-consistent $SU(3)$
chiral quark-soliton model \cite{Kim-Radii-soliton},  $\sqrt{\langle r^2\rangle_{D^+}}=0.43\,\rm{fm}$, $\sqrt{\langle r^2\rangle_{D^0}}=0.55\,\rm{fm}$ from the light-front quark model \cite{Hwang-Radii-LF}. In the present work, though we refer the color-singlet-color-singlet type pentaquark states as the pentaquark molecular states, they have the average spatial sizes as that of the typical heavy mesons and baryons, and are compact objects. For example, a loosely bound $\bar{D}^0\Sigma_c^+$ molecular state with the physical meson $\bar{D}^0$ and baryon $\Sigma_c^+$ should have average  spatial size $\sqrt{\langle r^2\rangle}\geq 1.36\,\rm{fm}$, which is too large to be interpolated  by the local currents.

The currents $J(0)$, $J_\mu(0)$ and $J_{\mu\nu}(0)$ couple potentially to the ${\frac{1}{2}}^\mp$, ${\frac{1}{2}}^\pm$, ${\frac{3}{2}}^\mp$ and ${\frac{1}{2}}^\mp$, ${\frac{3}{2}}^\pm$, ${\frac{5}{2}}^\mp$  hidden-charm  pentaquark molecular  states $P_{\frac{1}{2}}^{\mp}$, $P_{\frac{1}{2}}^{\pm}$, $P_{\frac{3}{2}}^{\mp}$ and $P_{\frac{1}{2}}^{\mp}$, $P_{\frac{3}{2}}^{\pm}$, $P_{\frac{5}{2}}^{\mp}$, respectively,
\begin{eqnarray}\label{J-lamda-1}
\langle 0| J (0)|P_{\frac{1}{2}}^{-}(p)\rangle &=&\lambda^{-}_{\frac{1}{2}} U^{-}(p,s) \, ,\nonumber  \\
\langle 0| J (0)|P_{\frac{1}{2}}^{+}(p)\rangle &=&\lambda^{+}_{\frac{1}{2}}i\gamma_5 U^{+}(p,s) \, ,
 \end{eqnarray}

 \begin{eqnarray}\label{J-lamda-2}
\langle 0| J_{\mu} (0)|P_{\frac{1}{2}}^{+}(p)\rangle &=&f^{+}_{\frac{1}{2}}p_\mu U^{+}(p,s) \, , \nonumber \\
\langle 0| J_{\mu} (0)|P_{\frac{1}{2}}^{-}(p)\rangle &=&f^{-}_{\frac{1}{2}}p_\mu i\gamma_5 U^{-}(p,s) \, , \nonumber\\
\langle 0| J_{\mu} (0)|P_{\frac{3}{2}}^{-}(p)\rangle &=&\lambda^{-}_{\frac{3}{2}} U^{-}_\mu(p,s) \, , \nonumber \\
\langle 0| J_{\mu} (0)|P_{\frac{3}{2}}^{+}(p)\rangle &=&\lambda^{+}_{\frac{3}{2}}i\gamma_5 U^{+}_{\mu}(p,s) \, ,
\end{eqnarray}

 \begin{eqnarray}\label{J-lamda-3}
\langle 0| J_{\mu\nu} (0)|P_{\frac{1}{2}}^{-}(p)\rangle &=&g^{-}_{\frac{1}{2}}p_\mu p_\nu U^{-}(p,s) \, , \nonumber\\
\langle 0| J_{\mu\nu} (0)|P_{\frac{1}{2}}^{+}(p)\rangle &=&g^{+}_{\frac{1}{2}}p_\mu p_\nu i\gamma_5 U^{+}(p,s) \, , \nonumber\\
\langle 0| J_{\mu\nu} (0)|P_{\frac{3}{2}}^{+}(p)\rangle &=&f^{+}_{\frac{3}{2}} \left[p_\mu U^{+}_{\nu}(p,s)+p_\nu U^{+}_{\mu}(p,s)\right] \, , \nonumber\\
\langle 0| J_{\mu\nu} (0)|P_{\frac{3}{2}}^{-}(p)\rangle &=&f^{-}_{\frac{3}{2}} i\gamma_5\left[p_\mu U^{-}_{\nu}(p,s)+p_\nu U^{-}_{\mu}(p,s)\right] \, , \nonumber\\
\langle 0| J_{\mu\nu} (0)|P_{\frac{5}{2}}^{-}(p)\rangle &=&\sqrt{2}\lambda^{-}_{\frac{5}{2}} U^{-}_{\mu\nu}(p,s) \, , \nonumber\\
\langle 0| J_{\mu\nu} (0)|P_{\frac{5}{2}}^{+}(p)\rangle &=&\sqrt{2}\lambda^{+}_{\frac{5}{2}}i\gamma_5 U^{+}_{\mu\nu}(p,s) \, ,
\end{eqnarray}
where the $U^\pm(p,s)$, $U^{\pm}_\mu(p,s)$ and $U^{\pm}_{\mu\nu}(p,s)$ are the Dirac and  Rarita-Schwinger spinors \cite{WZG-IJMPA-mole,Wang1508-EPJC,WangHuang1508-1,WangHuang1508-2,WangHuang1508-3,WZG-IJMPA-penta}.

 At the hadron side of the correlation functions $\Pi(p)$, $\Pi_{\mu\nu}(p)$ and $\Pi_{\mu\nu\alpha\beta}(p)$, we isolate the  ground state contributions from the hidden-charm pentaquark molecular states with the spin-parity $J^P={\frac{1}{2}}^\pm$, ${\frac{3}{2}}^\pm$ and ${\frac{5}{2}}^\pm$ respectively without contaminations according to the current-hadron couplings shown in Eqs.\eqref{J-lamda-1}-\eqref{J-lamda-3}, and get the hadronic representation \cite{WZG-IJMPA-mole,Wang1508-EPJC,WangHuang1508-1,WangHuang1508-2,WangHuang1508-3,WZG-IJMPA-penta},
\begin{eqnarray}
  \Pi(p) & = & {\lambda^{-}_{\frac{1}{2}}}^2  {\!\not\!{p}+ M_{-} \over M_{-}^{2}-p^{2}  } +  {\lambda^{+}_{\frac{1}{2}}}^2  {\!\not\!{p}- M_{+} \over M_{+}^{2}-p^{2}  } +\cdots  \, ,\nonumber\\
  &=&\Pi_{\frac{1}{2}}^1(p^2)\!\not\!{p}+\Pi_{\frac{1}{2}}^0(p^2)\, ,
\end{eqnarray}

 \begin{eqnarray}
   \Pi_{\mu\nu}(p) & = & {\lambda^{-}_{\frac{3}{2}}}^2  {\!\not\!{p}+ M_{-} \over M_{-}^{2}-p^{2}  } \left(- g_{\mu\nu}\right)+  {\lambda^{+}_{\frac{3}{2}}}^2  {\!\not\!{p}- M_{+} \over M_{+}^{2}-p^{2}  } \left(- g_{\mu\nu}\right)   +\cdots  \, ,\nonumber\\
   &=&-\Pi_{\frac{3}{2}}^1(p^2)\!\not\!{p}\,g_{\mu\nu}-\Pi_{\frac{3}{2}}^0(p^2)\,g_{\mu\nu}+\cdots\, ,
\end{eqnarray}

 \begin{eqnarray}
\Pi_{\mu\nu\alpha\beta}(p) & = & {\lambda^{-}_{\frac{5}{2}}}^2  {\!\not\!{p}+ M_{-} \over M_{-}^{2}-p^{2}  } \left( g_{\mu\alpha}g_{\nu\beta}+g_{\mu\beta}g_{\nu\alpha}
\right)+ {\lambda^{+}_{\frac{5}{2}}}^2  {\!\not\!{p}- M_{+} \over M_{+}^{2}-p^{2}  } \left( g_{\mu\alpha}g_{\nu\beta}+g_{\mu\beta}g_{\nu\alpha}\right) +\cdots \, , \nonumber\\
&=&\Pi_{\frac{5}{2}}^1(p^2)\!\not\!{p}\left( g_{\mu\alpha}g_{\nu\beta}+g_{\mu\beta}g_{\nu\alpha}\right)+\Pi_{\frac{5}{2}}^0(p^2)\,\left( g_{\mu\alpha}g_{\nu\beta}+g_{\mu\beta}g_{\nu\alpha}\right)+ \cdots \, .
\end{eqnarray}
There are other spinor structures, which are not shown explicitly, we choose the components corresponding to the spinor structures $\!\not\!{p}$,  $1$,   $\!\not\!{p}g_{\mu\nu}$, $g_{\mu\nu}$ and $\!\not\!{p}\left(g_{\mu\alpha}g_{\nu\beta}+g_{\mu\beta}g_{\nu\alpha}\right)$, $g_{\mu\alpha}g_{\nu\beta}+g_{\mu\beta}g_{\nu\alpha}$ in  the correlation functions $\Pi(p)$, $\Pi_{\mu\nu}(p)$ and $\Pi_{\mu\nu\alpha\beta}(p)$ respectively to investigate  the $J^P={\frac{1}{2}}^\mp$, ${\frac{3}{2}}^\mp$ and ${\frac{5}{2}}^\mp$ pentaquark molecular states.

Now we take a not long digression to discuss the isospins of the interpolating currents.  From Eqs.\eqref{current-J}-\eqref{current-Jmunu}, we can see clearly that the currents $J^{\bar{D}\Sigma_c}(x)$,  $J^{\bar{D}\Sigma_c^*}_{\mu}(x)$,  $ J^{\bar{D}^*\Sigma_c}_{\mu}(x)$ and $J^{\bar{D}^*\Sigma_c^*}_{\mu\nu}(x)$ without strangeness  have the same isospin structures, while the currents
 $J^{\bar{D}\Xi^\prime_c}(x)$, $J^{\bar{D}\Xi_c^*}_{\mu}(x)$,  $ J^{\bar{D}^*\Xi^\prime_c}_{\mu}(x)$ and $J^{\bar{D}^*\Xi_c^*}_{\mu\nu}(x)$ with strangeness also have the same isospin structures, and they can be transformed into each other with the simple replacements  $d \leftrightarrow s$. It is a good object to explore  the light flavor $SU(3)$ breaking effects.  We can rewrite the currents $J^{\bar{D}^0\Sigma_c^+}(x)$ and $J^{\bar{D}^0\Xi_c^{\prime+}}(x)$ in terms of the isospin eigenstates,
\begin{eqnarray}
J^{\bar{D}^0\Sigma_c^+}(x)&=&J_{\bar{D}^0}(x)J_{\Sigma_c^+}(x)=\frac{1}{\sqrt{3}}J_{\bar{D}\Sigma_c}^{\frac{1}{2}}(x)+\sqrt{\frac{2}{3}}J_{\bar{D}\Sigma_c}^{\frac{3}{2}}(x) \, ,\nonumber\\
J^{\bar{D}^0\Xi_c^{\prime+}}(x)&=&J_{\bar{D}^0}(x)J_{\Xi_c^{\prime+}}(x)=J_{\bar{D}\Xi^\prime_c}^{1}(x) \, ,
\end{eqnarray}
where
\begin{eqnarray}
J_{\bar{D}\Sigma_c}^{\frac{1}{2}}(x)&=&\frac{1}{\sqrt{3}}J_{\bar{D}^0}(x)J_{\Sigma_c^+}(x)-\sqrt{\frac{2}{3}}J_{\bar{D}^-}(x)J_{\Sigma_c^{++}}(x) \, ,\nonumber\\
J_{\bar{D}\Sigma_c}^{\frac{3}{2}}(x)&=&\sqrt{\frac{2}{3}}J_{\bar{D}^0}(x)J_{\Sigma_c^+}(x)+\frac{1}{\sqrt{3}}J_{\bar{D}^-}(x)J_{\Sigma_c^{++}}(x) \, ,
\end{eqnarray}
the $J_{\bar{D}^0}(x)$, $J_{\bar{D}^-}(x)$, $J_{\Sigma_c^+}(x)$, $J_{\Sigma_c^{++}}(x)$ and $J_{\Xi_c^{\prime+}}(x)$ are the standard currents for the mesons and baryons, respectively, the superscripts $\frac{1}{2}$, $1$ and $\frac{3}{2}$ stand for the isospins of the interpolating currents.
Now let us estimate the isospin breaking effects,
\begin{eqnarray}
 \langle0|T\left\{J_{\bar{D}^0\Sigma_c^+}(x)\bar{J}_{\bar{D}^0\Sigma_c^+}(0)\right\}|0\rangle &=& \frac{1}{3}\langle0|T\left\{J^{\frac{1}{2}}_{\bar{D}\Sigma_c}(x)\bar{J}^{\frac{1}{2}}_{\bar{D}\Sigma_c}(0)\right\}|0\rangle
 +\frac{2}{3}\langle0|T\left\{J^{\frac{3}{2}}_{\bar{D}\Sigma_c}(x)\bar{J}^{\frac{3}{2}}_{\bar{D}\Sigma_c}(0)\right\}|0\rangle\, , \nonumber\\
  \langle0|T\left\{J_{\bar{D}^{-}\Sigma_c^{++}}(x)\bar{J}_{\bar{D}^{-}\Sigma_c^{++}}(0)\right\}|0\rangle &=& \frac{2}{3}\langle0|T\left\{J^{\frac{1}{2}}_{\bar{D}\Sigma_c}(x)\bar{J}^{\frac{1}{2}}_{\bar{D}\Sigma_c}(0)\right\}|0\rangle
 +\frac{1}{3}\langle0|T\left\{J^{\frac{3}{2}}_{\bar{D}\Sigma_c}(x)\bar{J}^{\frac{3}{2}}_{\bar{D}\Sigma_c}(0)\right\}|0\rangle\, , \nonumber\\
\end{eqnarray}
where the current $J_{\bar{D}^-\Sigma_c^{++}}(x)=J_{\bar{D}^-}(x)J_{\Sigma_c^{++}}(x)$.
The isospin breaking effects between the vacuum matrix elements  $\langle0|T\left\{J_{\bar{D}^0\Sigma_c^+}(x)\bar{J}_{\bar{D}^0\Sigma_c^+}(0)\right\}|0\rangle$ and $\langle0|T\left\{J_{\bar{D}^{-}\Sigma_c^{++}}(x)\bar{J}_{\bar{D}^{-}\Sigma_c^{++}}(0)\right\}|0\rangle$ at the quark-gluon level are suppressed by a factor $\frac{1}{4N_c}$, which are of minor importance and can be neglected safely in the large $N_c$ limit. Then we can estimate (or obtain the conclusion tentatively) that the currents $J_{\bar{D}\Sigma_c}^{\frac{1}{2}}(x)$ and $J_{\bar{D}\Sigma_c}^{\frac{3}{2}}(x)$ couple potentially to the spin-$\frac{1}{2}$ $\bar{D}\Sigma_c$  pentaquark molecular states  with  almost  degenerated masses but different pole residues,  thereafter, we will not distinguish the isospins $I=\frac{1}{2}$ and $\frac{3}{2}$ as we are only interested in the molecule masses, just like in the previous works \cite{Penta-mole-QCDSR-Chen,Penta-mole-QCDSR-Zhang,Penta-mole-QCDSR-Azizi,WZG-IJMPA-mole}. Furthermore, the current $J_{\bar{D}\Xi^\prime_c}^{1}(x)$ has the quantum numbers $I=1$ and $I_3=1$,  we can add the superscripts $I_3$ to distinguish the components in the isospin triplet and singlet,
\begin{eqnarray}
J_{\bar{D}\Xi^\prime_c}^{1,1}(x) &= &J_{\bar{D}^0}(x)J_{\Xi_c^{\prime+}}(x)\, , \nonumber\\
J_{\bar{D}\Xi^\prime_c}^{1,0}(x) &= &\frac{1}{\sqrt{2}}J_{\bar{D}^0}(x)J_{\Xi_c^{\prime0}}(x)+\frac{1}{\sqrt{2}}J_{\bar{D}^-}(x)J_{\Xi_c^{\prime+}}(x)\, ,\nonumber\\
J_{\bar{D}\Xi^\prime_c}^{1,-1}(x) &= &J_{\bar{D}^-}(x)J_{\Xi_c^{\prime0}}(x)\, ,\nonumber\\
J_{\bar{D}\Xi^\prime_c}^{0,0}(x) &= &\frac{1}{\sqrt{2}}J_{\bar{D}^0}(x)J_{\Xi_c^{\prime0}}(x)-\frac{1}{\sqrt{2}}J_{\bar{D}^-}(x)J_{\Xi_c^{\prime+}}(x)\, ,
\end{eqnarray}
they couple potentially to the pentaquark molecular states with almost  degenerated masses, again the isospin breaking effects are suppressed  by the factor $\frac{1}{4N_c}$, thereafter, we will not distinguish the isospins $I=1$ and $0$.

Now let us go back to the correlation functions. It is straightforward to  obtain the spectral densities at hadron  side through dispersion relation,
\begin{eqnarray}
\frac{{\rm Im}\Pi_{j}^1(s)}{\pi}&=& {\lambda^{-}_{j}}^2 \delta\left(s-M_{-}^2\right)+{\lambda^{+}_{j}}^2 \delta\left(s-M_{+}^2\right) =\, \rho^1_{j,H}(s) \, , \\
\frac{{\rm Im}\Pi^0_{j}(s)}{\pi}&=&M_{-}{\lambda^{-}_{j}}^2 \delta\left(s-M_{-}^2\right)-M_{+}{\lambda^{+}_{j}}^2 \delta\left(s-M_{+}^2\right)
=\rho^0_{j,H}(s) \, ,
\end{eqnarray}
where $j=\frac{1}{2}$, $\frac{3}{2}$, $\frac{5}{2}$, we add the subscript $H$ to represent the hadron side,
then we introduce the  weight functions $\sqrt{s}\exp\left(-\frac{s}{T^2}\right)$ and $\exp\left(-\frac{s}{T^2}\right)$ to obtain the QCD sum rules at the hadron side,
\begin{eqnarray}
\int_{4m_c^2}^{s_0}ds \left[\sqrt{s}\rho^1_{j,H}(s)+\rho^0_{j,H}(s)\right]\exp\left( -\frac{s}{T^2}\right)
&=&2M_{-}{\lambda^{-}_{j}}^2\exp\left( -\frac{M_{-}^2}{T^2}\right) \, ,
\end{eqnarray}
where the $s_0$ are the continuum threshold parameters and the $T^2$ are the Borel parameters.

It is also straightforward to accomplish the operator product expansion in the deep Euclidian space-time. For technical details in performing the operator product expansion for the correlation functions in exploring the  multiquark states with hidden-charm, one can consults Refs.\cite{Wang1508-EPJC,WangHuang1508-1,WangHuang1508-2,WangHuang1508-3,Wang-tetraquark-QCDSR-1,Wang-tetraquark-QCDSR-2,
Wang-tetraquark-QCDSR-3,Wang-tetraquark-QCDSR-4,Wang-molecule-QCDSR-1,Wang-molecule-QCDSR-2}.   If we contract the quark fields in the correlation functions in Eq.\eqref{CF-Pi}
with the Wick's theorem, we can observe clearly that there are two heavy quark propagators and three light quark propagators, if each heavy quark line emits a gluon and each light quark line contributes  a quark-antiquark pair, we obtain a quark-gluon mixed operator $g_sG_{\mu\nu}g_sG_{\alpha\beta}\bar{q}q\bar{q}q\bar{q}q$ with $q=u$, $d$ or $s$, which is of dimension 13, and leads to the vacuum condensates $\langle\bar{q} q\rangle \langle\bar{q}g_s\sigma Gq\rangle^2 $ and $\langle \bar{q}q\rangle^3\langle \frac{\alpha_s}{\pi}GG\rangle$. It is better to take  account of the vacuum condensates up to dimension $n=13$ at least.
On the other hand, the quark-gluon operators can be counted by the fine structure constant $\alpha_s=\frac{g_s^2}{4\pi}$ with  the orders $\mathcal{O}( \alpha_s^{k})$,
where
$k=0$, $\frac{1}{2}$, $1$, $\frac{3}{2}$, $\cdots$.
In the  present work, we  take the truncations $n\leq 13$ and $k\leq 1$ in a consistent way,
the quark-gluon operators of the orders $\mathcal{O}( \alpha_s^{k})$ with $k\leq 1$ are given full considerations,
 while in previous work \cite{WZG-IJMPA-mole},
   we neglected the vacuum  condensates  $\langle \bar{q}q\rangle\langle \frac{\alpha_s}{\pi}GG\rangle$,
$\langle \bar{q}q\rangle^2\langle \frac{\alpha_s}{\pi}GG\rangle$ and $\langle \bar{q}q\rangle^3\langle \frac{\alpha_s}{\pi}GG\rangle$  due to their small contributions.
Furthermore, we take  account of the light flavor $SU(3)$ mass-breaking effects by including the contributions of the order $\mathcal{O}(m_s)$ in a consistent way.

Now we take a short digression to discuss the higher dimensional vacuum condensates. In the QED, we deal with the  perturbative vacuum, the vacuum expectation values of the normal-ordered  electron-photon operators can be set be zero, for example,  $\langle0|:\bar{e}e:|0\rangle=0$, $\langle0|:\bar{e}\sigma \cdot F e:|0\rangle=0$,
 $\langle0|:\bar{e}e\bar{e}e:|0\rangle=0$, etc.

 In the QCD, we deal with the non-perturbative vacuum, and have to resort to non-zero vacuum expectation values of the normal-ordered  quark-gluon operators to describe the hadron properties in a satisfactory way, for example, $\langle0|:\bar{q}_{\alpha}^{i} q_{\beta}^{j}:|0\rangle\neq0$, $\langle0|:\bar{q}_{\alpha}^{i}   q_{\beta}^{j}g_sG^a_{\mu\nu}:|0\rangle\neq0$, $\langle0|:\bar{q}_{\alpha}^{i} q_{\beta}^{j}\bar{q}_{\lambda}^{m} q_{\tau}^{n}:|0\rangle\neq0$, etc, where the $i$, $j$, $m$ and $n$ are color indexes, the $\alpha$, $\beta$, $\lambda$ and $\tau$ are Dirac spinor indexes. We usually parameterize the vacuum matrix elements  in terms of
 $\langle0|:\bar{q}_{\alpha}^{i} q_{\beta}^{j}:|0\rangle=\frac{1}{12}\langle0|:\bar{q} q:|0\rangle \delta_{ij}\delta_{\alpha\beta}$, $\langle0|:\bar{q}_{\alpha}^{i}  q_{\beta}^{j}G^a_{\mu\nu}:|0\rangle=\frac{1}{192}\langle0|:\bar{q} g_s\sigma\cdot G q:|0\rangle\left(\sigma_{\mu\nu}\right)_{\beta\alpha}\frac{\lambda^a_{ji}}{2}$, $\langle0|:\bar{q}_{\alpha}^{i} q_{\beta}^{j}\bar{q}_{\lambda}^{m} q_{\tau}^{n}:|0\rangle=\frac{\varrho}{144}\langle0|:\bar{q} q:|0\rangle^2 \left(\delta_{ij}\delta_{mn}\delta_{\alpha\beta}\delta_{\lambda\tau}-\delta_{in}\delta_{jm}\delta_{\alpha\tau}\delta_{\beta\lambda} \right)$, or
 $\frac{1}{144}\langle0|:\bar{q} q\bar{q} q:|0\rangle \left(\delta_{ij}\delta_{mn}\delta_{\alpha\beta}\delta_{\lambda\tau}-\delta_{in}\delta_{jm}\delta_{\alpha\tau}\delta_{\beta\lambda} \right)$, etc, where the $\lambda^a$ are the Gell-mann matrixes.
 Except for the
 quark condensates, which indicate  spontaneous breaking of the Chiral symmetry through the Gell-Mann-Oakes-Renner relation
 $f^2_{\pi}m^2_{\pi}=-2(m_u+m_d)\langle0|:\bar{q} q:|0\rangle$ \cite{GMOR}, other vacuum condensates, such as $\langle0|:\bar{q} g_s\sigma\cdot G q:|0\rangle$,
 $\langle0|:\bar{q} q\bar{q} q:|0\rangle$, $\varrho\langle0|:\bar{q} q:|0\rangle^2$, $\cdots$ are just parameters introduced by hand to describe the non-pertubative vacuum.

 We can parameterize the non-perturbative properties  in one way or the other, then confront them to the experimental data on the multiquark states to obtain the optimal values.  In the QCD sum rules for the multiquark states, the $\varrho\langle\bar{q}q\rangle^2$ play  an important role, and influence  the convergent behaviors of the operator product expansion and the pole contributions remarkably, therefore influence  the predictions remarkably,
 large values of the $\varrho$ maybe destroy the platforms \cite{WZG-DvDvDv}, for example, in the present case, if we take the value $\varrho=2(3)$ in the QCD sum rules for the $\bar{D}\Sigma_c$ pentaquark molecular state, we can obtain the uncertainty $\delta M_{P}=-0.10(-0.16)\,\rm{GeV}$, which is of the same order of the total uncertainty from other parameters, and a very bad  platform (in other words, no platform at all).  In calculations, we observe that the optimal value is $\varrho=1$,  vacuum saturation (factorizaiton) works well in the QCD sum rules for the multiquark states \cite{Wang1508-EPJC,WangHuang1508-1,WangHuang1508-2,WangHuang1508-3,WZG-IJMPA-penta,WZG-Pcs4459,Wang-tetraquark-QCDSR-1,Wang-tetraquark-QCDSR-2,
 Wang-tetraquark-QCDSR-3,Wang-tetraquark-QCDSR-4,Wang-molecule-QCDSR-1,Wang-molecule-QCDSR-2,WZG-DvDvDv}.

 In the QCD sum rules for the $q\bar{q}$, $q\bar{Q}$, $Q\bar{Q}$  mesons,  the $\varrho\langle\bar{q}q\rangle^2$  are always companied with the fine-structure constant $\alpha_s$, and play a  tiny role,
the deviation from  vacuum saturation (factorizaiton) $\varrho=1$, for example, $\varrho=2\sim 3$, cannot make much difference in the numerical predictions,
although in some cases the values  $\varrho>1$ can lead to better QCD sum rules  \cite{Review-rho-varrho,Narison-rho}.

 Once the corresponding analytical spectral densities $\rho^1_{j,QCD}(s)$ and $\rho^0_{j,QCD}(s)$ at the quark-gluon level are obtained,  we can take the
quark-hadron duality below the continuum thresholds  $s_0$ and introduce the weight functions $\sqrt{s}\exp\left(-\frac{s}{T^2}\right)$ and $\exp\left(-\frac{s}{T^2}\right)$ to obtain  the  QCD sum rules:
\begin{eqnarray}\label{QCDN}
2M_{-}{\lambda^{-}_{j}}^2\exp\left( -\frac{M_{-}^2}{T^2}\right)
&=& \int_{4m_c^2}^{s_0}ds \left[\sqrt{s}\rho^1_{j,QCD}(s)+\rho^0_{j,QCD}(s)\right]\exp\left( -\frac{s}{T^2}\right)\, ,
\end{eqnarray}
the explicit expressions of the  spectral densities $\rho^1_{j,QCD}(s)$ and $\rho^0_{j,QCD}(s)$ at the quark level  are neglected for simplicity.

We differentiate   Eq.\eqref{QCDN} with respect to  $\tau=\frac{1}{T^2}$, then eliminate the
 pole residues $\lambda^{-}_{j}$ with $j=\frac{1}{2}$, $\frac{3}{2}$, $\frac{5}{2}$ to  obtain the QCD sum rules for
 the masses of the pentaquark molecular  states,
 \begin{eqnarray}\label{QCDSR-M}
 M^2_{-} &=& \frac{-\frac{d}{d \tau}\int_{4m_c^2}^{s_0}ds \,\left[\sqrt{s}\,\rho^1_{QCD}(s)+\,\rho^0_{QCD}(s)\right]\exp\left(- \tau s\right)}{\int_{4m_c^2}^{s_0}ds \left[\sqrt{s}\,\rho_{QCD}^1(s)+\,\rho^0_{QCD}(s)\right]\exp\left( -\tau s\right)}\, ,
 \end{eqnarray}
where the spectral densities $\rho_{QCD}^1(s)=\rho_{j,QCD}^1(s)$ and $\rho^0_{QCD}(s)=\rho^0_{j,QCD}(s)$.

\section{Numerical results and discussions}
We take  the standard values of the  vacuum condensates
$\langle\bar{q}q \rangle=-(0.24\pm 0.01\, \rm{GeV})^3$,  $\langle\bar{s}s \rangle=(0.8\pm0.1)\langle\bar{q}q \rangle$,
 $\langle\bar{q}g_s\sigma G q \rangle=m_0^2\langle \bar{q}q \rangle$, $\langle\bar{s}g_s\sigma G s \rangle=m_0^2\langle \bar{s}s \rangle$,
$m_0^2=(0.8 \pm 0.1)\,\rm{GeV}^2$, $\langle \frac{\alpha_s
GG}{\pi}\rangle=0.012\pm0.004\,\rm{GeV}^4$    at the energy scale  $\mu=1\, \rm{GeV}$
\cite{SVZ79,PRT85,ColangeloReview}, and  take the $\overline{MS}$ masses $m_{c}(m_c)=(1.275\pm0.025)\,\rm{GeV}$
 and $m_s(\mu=2\,\rm{GeV})=(0.095\pm0.005)\,\rm{GeV}$
 from the Particle Data Group \cite{PDG}.
Furthermore,  we take account of
the energy-scale dependence of  the quark condensates, mixed quark condensates and $\overline{MS}$ masses according to  the renormalization group equation \cite{Narison-mix},
 \begin{eqnarray}
 \langle\bar{q}q \rangle(\mu)&=&\langle\bar{q}q\rangle({\rm 1 GeV})\left[\frac{\alpha_{s}({\rm 1 GeV})}{\alpha_{s}(\mu)}\right]^{\frac{12}{33-2n_f}}\, , \nonumber\\
 \langle\bar{s}s \rangle(\mu)&=&\langle\bar{s}s \rangle({\rm 1 GeV})\left[\frac{\alpha_{s}({\rm 1 GeV})}{\alpha_{s}(\mu)}\right]^{\frac{12}{33-2n_f}}\, , \nonumber\\
 \langle\bar{q}g_s \sigma Gq \rangle(\mu)&=&\langle\bar{q}g_s \sigma Gq \rangle({\rm 1 GeV})\left[\frac{\alpha_{s}({\rm 1 GeV})}{\alpha_{s}(\mu)}\right]^{\frac{2}{33-2n_f}}\, ,\nonumber\\
  \langle\bar{s}g_s \sigma Gs \rangle(\mu)&=&\langle\bar{s}g_s \sigma Gs \rangle({\rm 1 GeV})\left[\frac{\alpha_{s}({\rm 1 GeV})}{\alpha_{s}(\mu)}\right]^{\frac{2}{33-2n_f}}\, ,\nonumber\\
m_c(\mu)&=&m_c(m_c)\left[\frac{\alpha_{s}(\mu)}{\alpha_{s}(m_c)}\right]^{\frac{12}{33-2n_f}} \, ,\nonumber\\
m_s(\mu)&=&m_s({\rm 2GeV} )\left[\frac{\alpha_{s}(\mu)}{\alpha_{s}({\rm 2GeV})}\right]^{\frac{12}{33-2n_f}}\, ,\nonumber\\
\alpha_s(\mu)&=&\frac{1}{b_0t}\left[1-\frac{b_1}{b_0^2}\frac{\log t}{t} +\frac{b_1^2(\log^2{t}-\log{t}-1)+b_0b_2}{b_0^4t^2}\right]\, ,
\end{eqnarray}
  where $t=\log \frac{\mu^2}{\Lambda^2}$, $b_0=\frac{33-2n_f}{12\pi}$, $b_1=\frac{153-19n_f}{24\pi^2}$, $b_2=\frac{2857-\frac{5033}{9}n_f+\frac{325}{27}n_f^2}{128\pi^3}$,  $\Lambda=213\,\rm{MeV}$, $296\,\rm{MeV}$  and  $339\,\rm{MeV}$ for the flavors  $n_f=5$, $4$ and $3$, respectively  \cite{PDG,Narison-mix}.

In the present work, we investigate   the hidden-charm pentaquark molecular  states with strangeness and without strangeness,  it is better to choose the flavor numbers $n_f=4$, and evolve all the input parameters to typical or special  energy scales  $\mu$, which satisfy the energy scale formula or modified energy scale formula \cite{Wang1508-EPJC,WangHuang1508-1,WangHuang1508-2,WangHuang1508-3,Wang-tetraquark-QCDSR-1,Wang-tetraquark-QCDSR-2,Wang-tetraquark-QCDSR-3,Wang-tetraquark-QCDSR-4,
Wang-molecule-QCDSR-1,Wang-molecule-QCDSR-2} with the updated value of the effective (or constituent) charmed quark mass ${\mathbb{M}}_c=1.85\,\rm{GeV}$ \cite{Wang-CPC-4390}. By comparing with the constituent quark masses based on analysis of the $J/\psi$ and $\Upsilon$
mass spectrum with the famous   Cornell potential \cite{Cornell}, we  introduce an uncertainty  ${\mathbb{M}}_c=1.85\pm0.01\,\rm{GeV}$.
Furthermore, we take  account of  the light flavor $SU(3)$ mass breaking effects, and prefer the modified energy scale formula $\mu=\sqrt{M^2_{X/Y/Z/P}-(2{\mathbb{M}}_c)^2}-k\,\mathbb{M}_s$ with the  $s$-quark numbers $k=0$, $1$, $2$, $3$ and the effective $s$-quark mass $\mathbb{M}_s=0.2\,\rm{GeV}$, which was   proved work well \cite{WZG-hidden-charm-mole}. Compared to the constituent quark mass, the effective $s$-quark mass $\mathbb{M}_s=0.2\,\rm{GeV}$ seems too small, as the effective $u/d$-quark masses $\mathbb{M}_{u/d}$ serve as a milestone and have been absorbed into the energy scales  $\mu$, the value $\mathbb{M}_s=0.2\,\rm{GeV}$ embodies the net $SU(3)$ mass-breaking effects.

We can rewrite the energy scale formula in the form,
\begin{eqnarray}\label{formula-Regge}
M^2_{X/Y/Z/P}&=&(\mu+k\,\delta)^2+{\rm Constants}\, ,
\end{eqnarray}
where the light flavor $SU(3)$ mass-breaking effects $\delta$ have the value $\mathbb{M}_s$,  the Constants have the value $4{\mathbb{M}}_c^2$, they are all fitted by the QCD sum rules. The $\mu$ and $\mathbb{M}_s$ embody the light degrees of freedom,  while the $4{\mathbb{M}}_c^2$ embodies the heavy degrees of freedom, the hidden-charm tetraquark and pentaquark (molecular) states can be divided into both the heavy and light degrees of freedom \cite{Wang1508-EPJC,WangHuang1508-1,WangHuang1508-2,WangHuang1508-3,Wang-tetraquark-QCDSR-2,Wang-tetraquark-QCDSR-3,Wang-tetraquark-QCDSR-4,
Wang-molecule-QCDSR-1,Wang-molecule-QCDSR-2}. The predicted tetraquark and pentaquark (molecular) masses and the pertinent  energy scales of the QCD spectral densities have a  Regge-trajectory-like relation \cite{WZG-DvDvDv}.

In Ref.\cite{WZG-IJMPA-mole}, we explore  the $\bar{D}\Sigma_c$, $\bar{D}\Sigma_c^*$, $\bar{D}^{*}\Sigma_c$ and
$  \bar{D}^{*}\Sigma_c^*$  pentaquark molecular states  with the QCD sum rules  at length, and obtain the conclusion that the energy scale formula $\mu=\sqrt{M^2_{X/Y/Z/P}-(2{\mathbb{M}}_c)^2}$ can enhance the pole contributions at the hadron side remarkably and improve the convergent behaviors of the operator product expansion remarkably. In fact, we take the energy scale formula as a constraint on the predicted molecule masses, which should be obeyed  in the QCD sum rules.
 In the present work, we consider the light flavor $SU(3)$ mass-breaking effects and  resort to the modified energy scale formula $\mu=\sqrt{M^2_{X/Y/Z/P}-(2{\mathbb{M}}_c)^2}-k\,\mathbb{M}_s$ to choose the best energy scales of the spectral densities at the quark-gluon level, and search for the best Borel parameters and continuum threshold parameters to satisfy the two fundamental criteria of the QCD sum rules, i.e. pole dominance at the hadron side and convergence of the operator product expansion at the QCD side, via trial and error.

Then we obtain  the Borel parameters, continuum threshold parameters $s_0$, optimal  energy scales of the spectral densities at the quark-gluon  level, and pole contributions of the ground state pentaquark molecular states, which are shown  plainly in Table \ref{Borel-pole}. From the table, we can see clearly that the contributions from the ground states are about or larger than $(40-60)\%$, the pole dominance criterion  is satisfied very well.
For the conventional hadrons, the QCD spectral densities $\rho(s)\sim s^n$ with $n\leq 1$ and $2$ for the mesons and baryons, respectively,
it is easy to satisfy the pole dominance criterion, as the integral,
\begin{eqnarray}
\int_{\Delta^2}^{s_0} ds s^{n} \exp\left( -\frac{s}{T^2}\right)\, ,
\end{eqnarray}
converges quickly even if we choose a large Borel parameter $T^2$, where the $\Delta^2$ is the threshold,  the uncertainty originates from the continuum threshold parameter $s_0$ is small.
For the multiquark  states, the QCD spectral densities $\rho(s)\sim s^n$ with $n\leq 4$ and $5$ for the tetraquark and pentaquark (molecular) states, respectively,
it is very difficult to satisfy the pole dominance criterion, as the integral, 
\begin{eqnarray}
\int_{\Delta^2}^{s_0} ds s^{n} \exp\left( -\frac{s}{T^2}\right)\, ,
\end{eqnarray}
converges very slowly even if we choose a rather small Borel parameter $T^2$. In general, we expect to choose $T^2=\mathcal{O}(M^2)$, the integral (or continuum state) is suppressed by a
fact $\exp\left( -\frac{s}{T^2}\right)\sim \exp\left( -\frac{M^2}{T^2}\right)\sim e^{-1}$. Thus, for the multiquark states, we have to resort to a much stringent suppression
of the continuum states, $T^2\ll M^2$. One may wonder that such a small Borel parameter might  lead to a bad convergent behavior in the operator product expansion.

\begin{figure}
 \centering
  \includegraphics[totalheight=5cm,width=7cm]{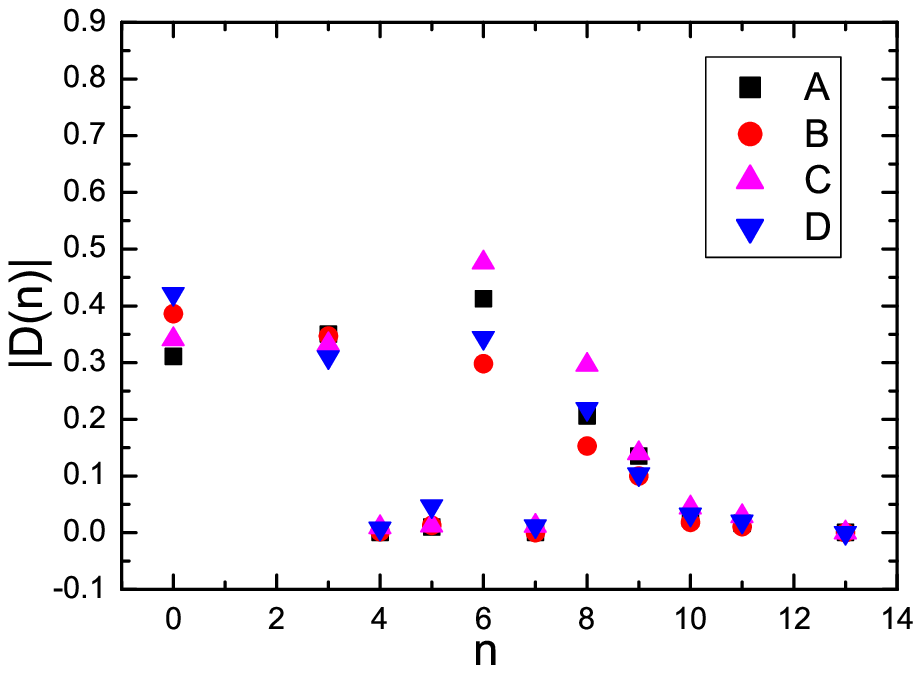}
  \includegraphics[totalheight=5cm,width=7cm]{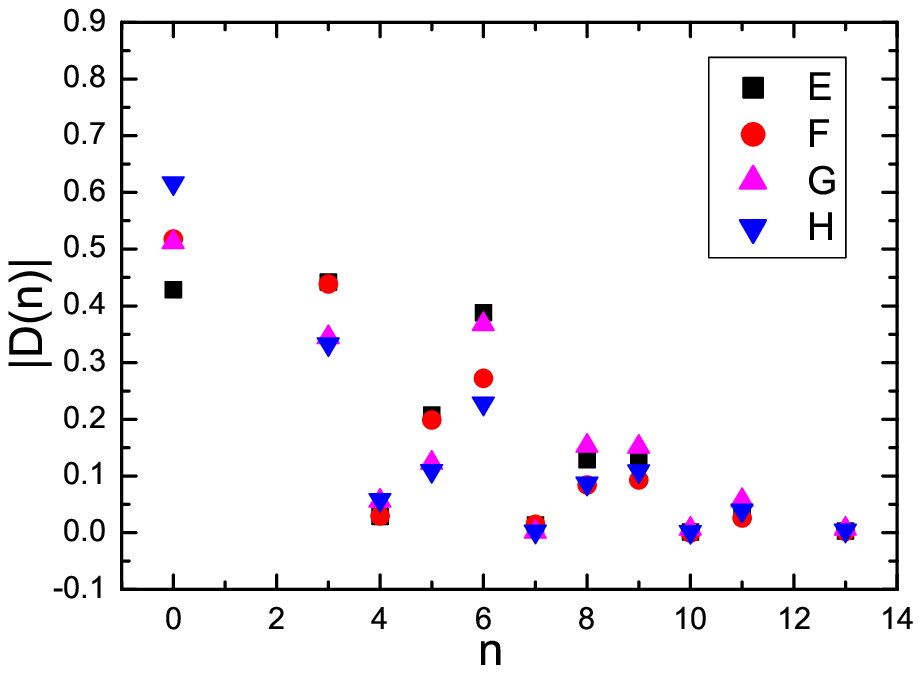}
  \caption{ The absolute values of the contributions of the vacuum condensates of dimension $n$,  where  the $A$, $B$, $C$, $D$, $E$, $F$, $G$ and $H$  denote the pentaquark molecular  states  $\bar{D}\Sigma_c$, $\bar{D}\Xi^\prime_c$, $\bar{D}\Sigma_c^*$, $\bar{D}\Xi_c^*$, $\bar{D}^{*}\Sigma_c$, $\bar{D}^{*}\Xi_c^\prime$,  $ \bar{D}^{*}\Sigma_c^*$ and $\bar{D}^{*}\Xi_c^*$, respectively.   }\label{OPE}
\end{figure}

In Fig.\ref{OPE}, we plot the absolute values of the $D(n)$ for the central values of the input parameters shown in Table \ref{Borel-pole}, where the $D(n)$ are defined by
\begin{eqnarray}\label{Dn}
D(n)&=& \frac{  \int_{4m_c^2}^{s_0} ds\,\rho_{n}(s)\,\exp\left(-\frac{s}{T^2}\right)}{\int_{4m_c^2}^{s_0} ds \,\rho(s)\,\exp\left(-\frac{s}{T^2}\right)}\, ,
\end{eqnarray}
 the $\rho_{n}(s)$ are the QCD spectral densities for the vacuum condensates of dimension $n$, and the total spectral densities
$\rho(s)=\sqrt{s}\,\rho^1_{QCD}(s)+\,\rho^0_{QCD}(s)$.
 From the figure, we can see  clearly  that although the largest contributions do not come from the terms $D(0)$ in some cases,
the vacuum condensates $\langle\bar{q}q\rangle\langle\bar{q}q\rangle$ and $\langle\bar{q}q\rangle\langle\bar{s}s\rangle$ with the dimension $6$ serve   as a milestone,   the absolute values of the  contributions $|D(n)|$ with $n\geq 6$ decrease  monotonically  and quickly with the increase of the dimensions $n$, the value $|D(13)|\approx 0$, the operator product expansion  converges very well. The two basic criteria of the QCD sum rules are all satisfied.

In calculations, we observe that the predicted molecule masses increase  monotonically  and slowly with the increase of the continuum threshold parameters $s_0$ if we fix the Borel parameter $T^2$;  on the other hand, larger continuum threshold parameters mean larger pole contributions. We truncate the continuum threshold parameters $s_0$ by requiring about the same pole contributions $(40-60)\%$ in all the QCD sum rules so as to reduce the uncertainties originate from the continuum threshold parameters $s_0$.

In previous work \cite{WZG-IJMPA-mole},
   we neglected the vacuum  condensates  $\langle \bar{q}q\rangle\langle \frac{\alpha_s}{\pi}GG\rangle$,
$\langle \bar{q}q\rangle^2\langle \frac{\alpha_s}{\pi}GG\rangle$ and $\langle \bar{q}q\rangle^3\langle \frac{\alpha_s}{\pi}GG\rangle$, which are of dimension $7$, $10$ and $13$, respectively,  due to their small contributions. From Fig.\ref{OPE}, we can see clearly that the vacuum condensates of $7$, $10$ and $13$ play a tiny role in the Borel windows indeed. We prefer to take account of those contributions because they lead to slightly larger pole contributions, therefore more reliable QCD sum rules. In the present work, we intend to explore the $SU(3)$-breaking effects, it is better to take account of those contributions in a consistent treatment.

\begin{figure}
 \centering
  \includegraphics[totalheight=5cm,width=7cm]{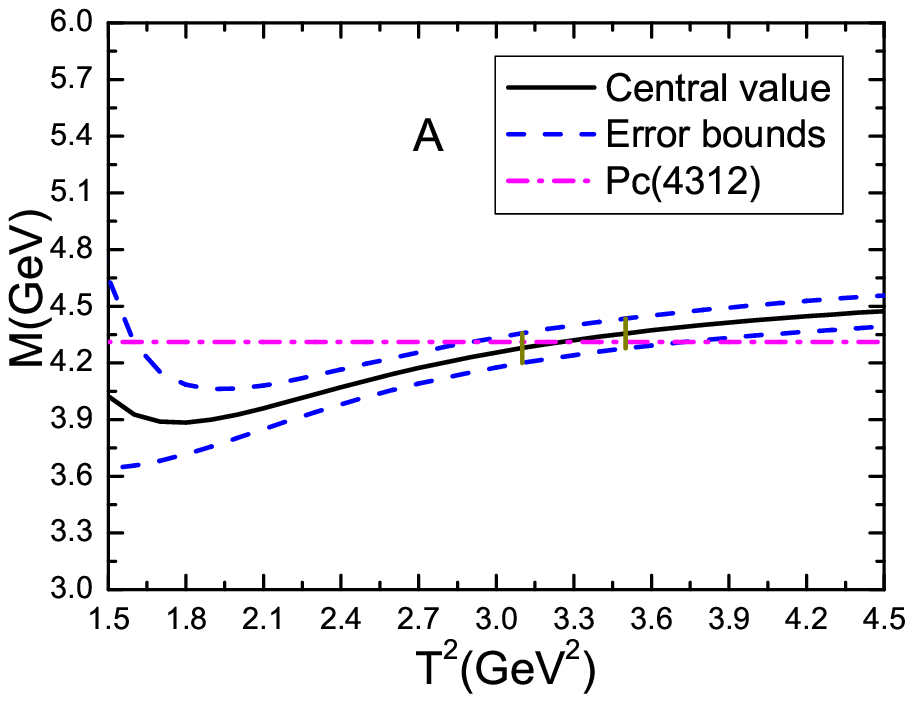}
   \includegraphics[totalheight=5cm,width=7cm]{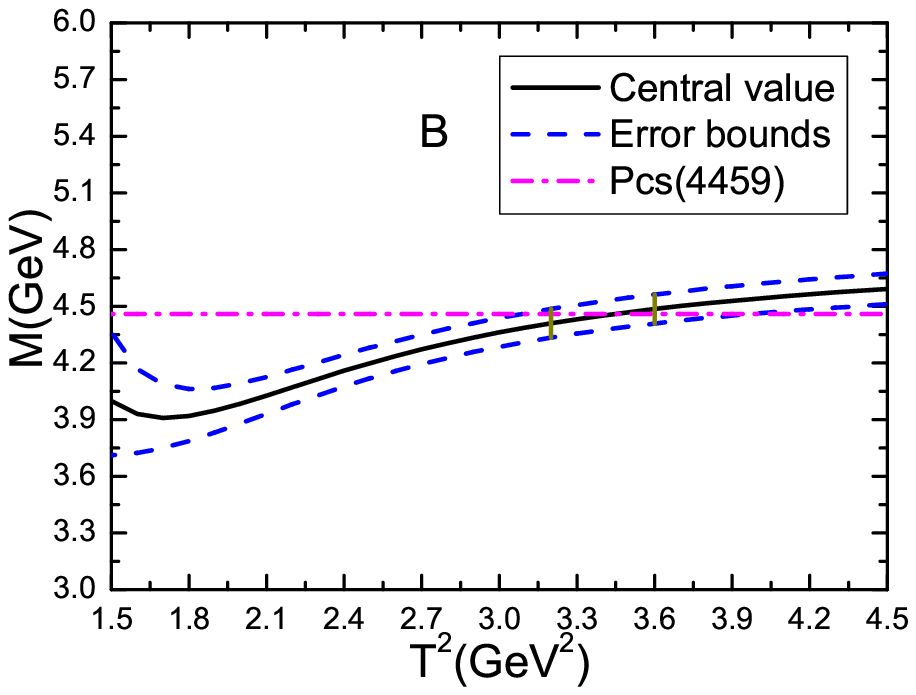}
   \includegraphics[totalheight=5cm,width=7cm]{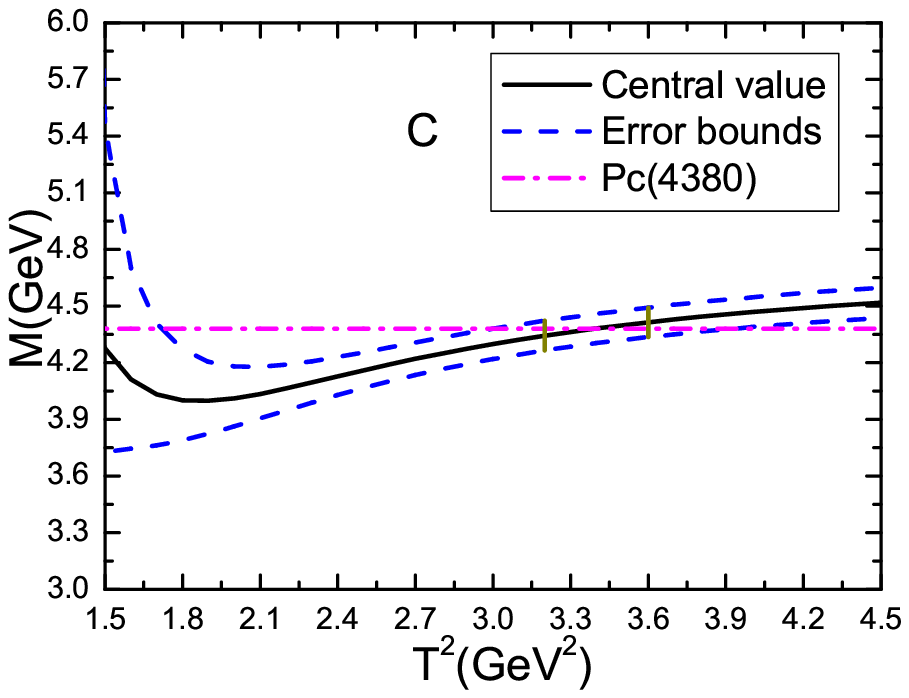}
  \includegraphics[totalheight=5cm,width=7cm]{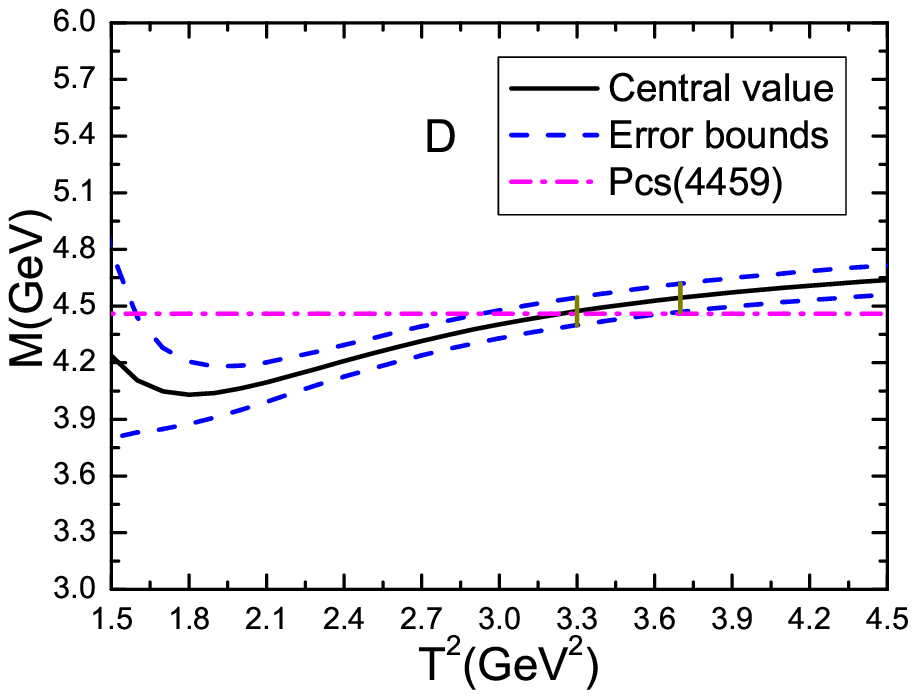}
   \includegraphics[totalheight=5cm,width=7cm]{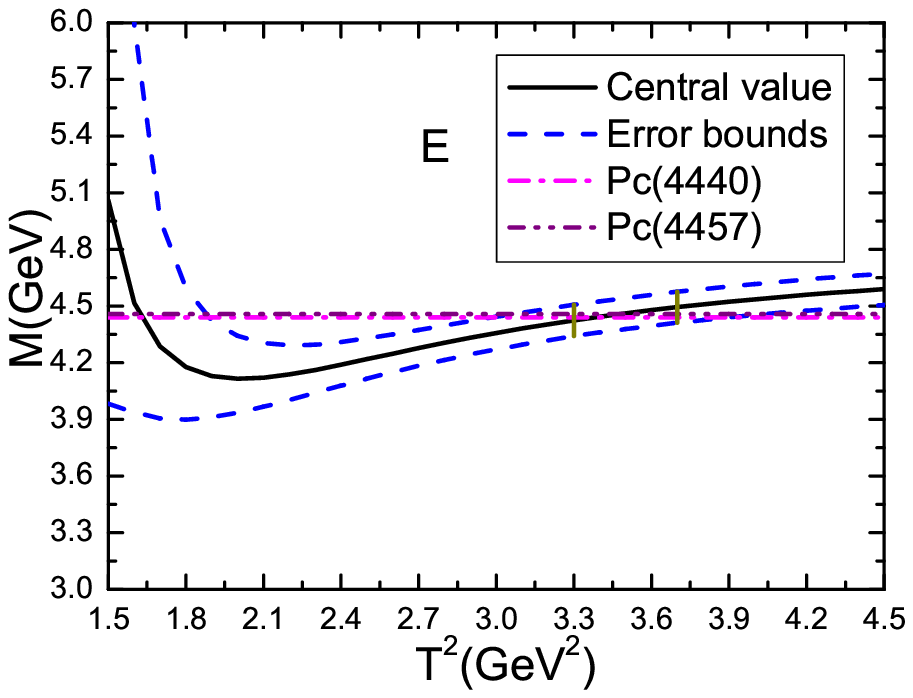}
   \includegraphics[totalheight=5cm,width=7cm]{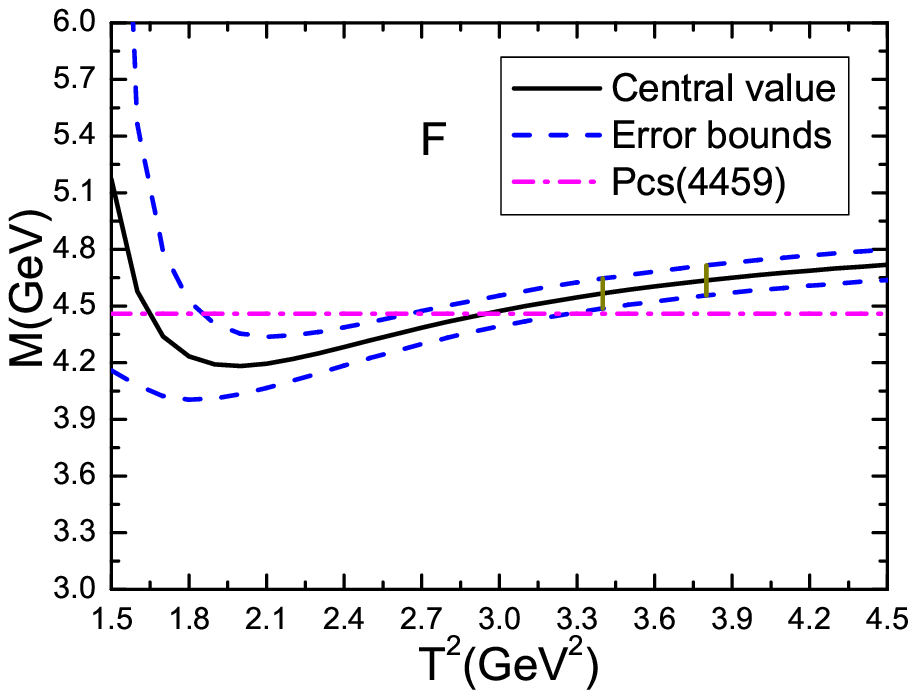}
    \includegraphics[totalheight=5cm,width=7cm]{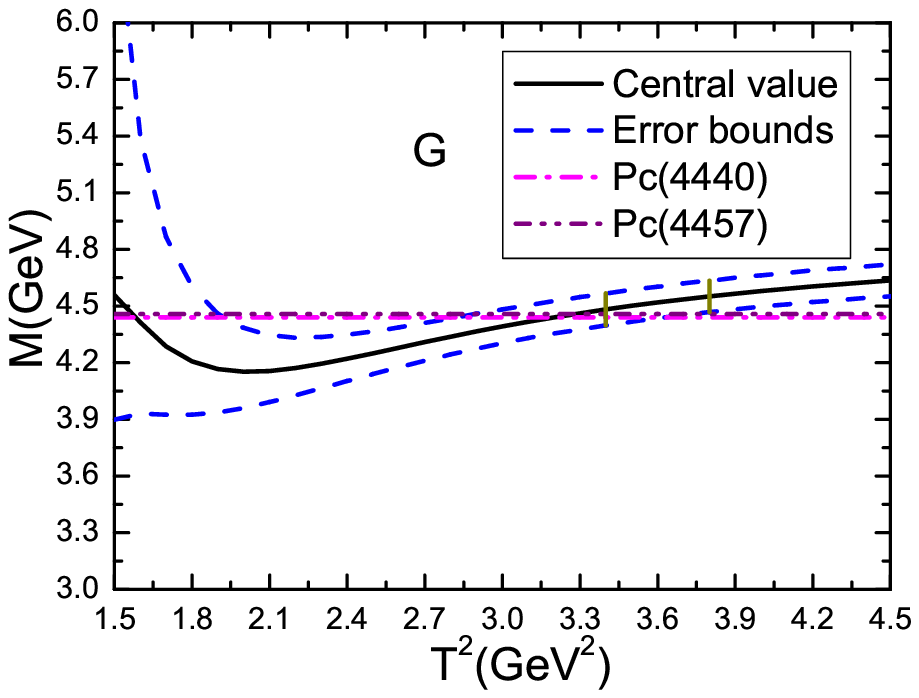}
  \includegraphics[totalheight=5cm,width=7cm]{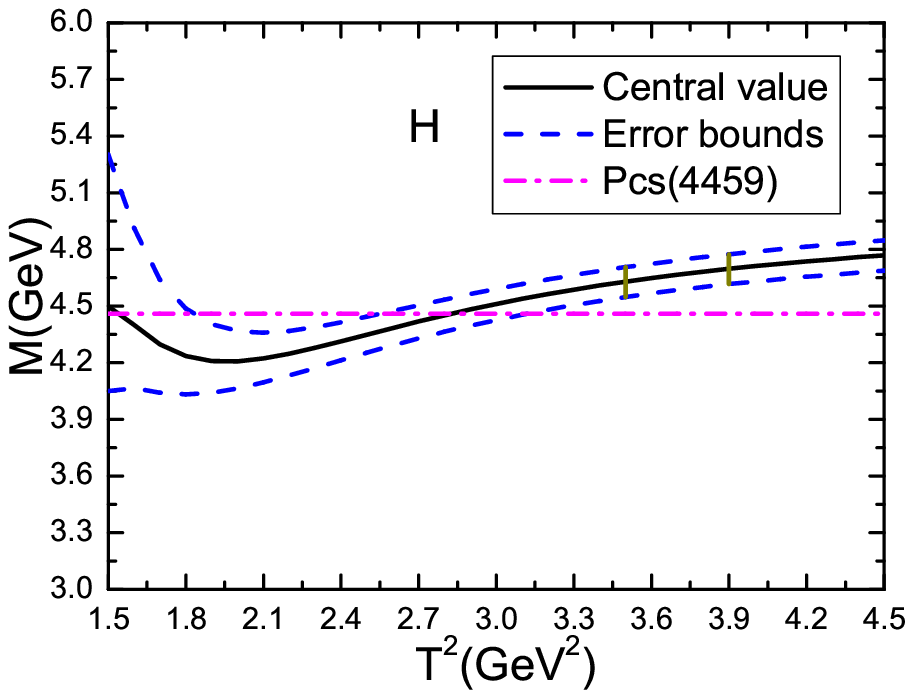}
  \caption{ The masses of the pentaquark molecular states  with variations of the Borel parameter $T^2$,  where  the $A$, $B$, $C$, $D$, $E$, $F$, $G$ and $H$  denote the pentaquark molecular  states  $\bar{D}\Sigma_c$, $\bar{D}\Xi^\prime_c$, $\bar{D}\Sigma_c^*$, $\bar{D}\Xi_c^*$, $\bar{D}^{*}\Sigma_c$, $\bar{D}^{*}\Xi_c^\prime$,  $ \bar{D}^{*}\Sigma_c^*$ and $\bar{D}^{*}\Xi_c^*$, respectively.   }\label{massDSigma-Borel}
\end{figure}

Finally we take  account of all uncertainties  of the input   parameters,
and obtain  the masses and pole residues of
 the $J^P={1\over 2}^{-}$, ${\frac{3}{2}}^-$  and ${\frac{5}{2}}^-$  hidden-charm pentaquark molecular states without  strangeness and with strangeness, which are shown explicitly in Table \ref{mass-pole-residue-tab} and Fig.\ref{massDSigma-Borel}. In Fig.\ref{massDSigma-Borel}, we plot the predicted masses of the hidden-charm pentaquark molecules without  strangeness and with strangeness according to variations of the Borel parameters, where the regions between the two short vertical lines are the Borel windows. From the figure, we can see clearly that there appear rather flat platforms in the Borel windows, the uncertainties come from the Borel parameters are rather small, which are compatible with the fact that the Borel parameters are just supplementary parameters, not physical quantities.   Furthermore, in the figure, we also present the experimental values of the masses of the $P_c(4312)$, $P_c(4380)$, $P_c(4440)$, $P_c(4457)$ and $P_{cs}(4459)$ from the LHCb collaboration \cite{LHCb-4380,LHCb-Pc4312,LHCb-Pcs4459-2012}.

The pentaquark (molecule) candidates $P_c(4312)$, $P_c(4380)$, $P_c(4440)$ and $P_c(4457)$ are observed in the $J/\psi p$ mass spectrum, their isospins are $I=\frac{1}{2}$, while  pentaquark (molecule) candidate $P_{cs}(4459)$ is observed in the $J/\psi \Lambda$ mass spectrum, its isospin is $I=0$.
 The present calculations  support assigning the $P_c(4312)$ as the $\bar{D}\Sigma_c$ pentaquark molecular state with the quantum numbers  $J^P={\frac{1}{2}}^-$ and $I=\frac{1}{2}$, assigning the $P_c(4380)$ as the $\bar{D}\Sigma_c^*$ pentaquark molecular state with the quantum numbers $J^P={\frac{3}{2}}^-$  and $I=\frac{1}{2}$,  assigning the $P_c(4440/4457)$ as the $\bar{D}^{*}\Sigma_c$ pentaquark molecular state with the quantum numbers $J^P={\frac{3}{2}}^-$  and $I=\frac{1}{2}$,   assigning the $P_{cs}(4459)$ as the $\bar{D}\Xi^\prime_c$ pentaquark molecular state with the quantum numbers $J^P={\frac{1}{2}}^-$  and $I=0$. However, we cannot
  exclude the possibilities of assigning the $P_c(4457)$ as the $\bar{D}^{*}\Sigma_c^*$ pentaquark molecular state with the quantum numbers $J^P={\frac{5}{2}}^-$  and $I=\frac{1}{2}$ and
  assigning the $P_{cs}(4459)$ as the $\bar{D}\Xi^*_c$ pentaquark molecular state with the quantum numbers $J^P={\frac{3}{2}}^-$  and $I=0$ due to the uncertainties, see Table \ref{mass-pole-residue-tab} and Fig.\ref{massDSigma-Borel}.
For example, it is marginal to assign the $P_c(4457)$ as the $\bar{D}^{*}\Sigma_c^*$ pentaquark molecular state with the quantum numbers $J^P={\frac{5}{2}}^-$  and $I=\frac{1}{2}$, as the $P_c(4457)$ lies at the bottom of the predicted mass of the $\bar{D}^{*}\Sigma_c^*$ pentaquark molecular state, see Fig.\ref{massDSigma-Borel}-G.

 From Tables \ref{Borel-pole}-\ref{mass-pole-residue-tab}, we can see that the modified energy scale formula  $\mu=\sqrt{M^2_{X/Y/Z/P}-(2{\mathbb{M}}_c)^2}-k\,\mathbb{M}_s$    with the  $s$-quark numbers $k=0$, $1$, $2$, $3$ and the effective $s$-quark mass $\mathbb{M}_s=0.2\,\rm{GeV}$ is satisfied very well  \cite{WZG-hidden-charm-mole}. On the other hand, the predicted masses for the pentaquark molecular states without strangeness and with strangeness have the relation, $M_{P_{cs}}-M_{P_c}\approx m_s-m_q\approx 0.13\sim0.15\,\rm{GeV}$, which is consistent with the light-flavor $SU(3)$ breaking effects for the heavy baryons in the flavor sextet ${\bf 6}_f$, $M_{\Xi^\prime_c}-M_{\Sigma_c}\approx M_{\Xi^*_c}-M_{\Sigma_c^*}\approx m_s-m_q\approx 0.13\,\rm{GeV}$ from the Particle Data Group \cite{PDG}.

  The present calculations indicate that there maybe exist the $\bar{D}\Sigma_c$ ($\bar{D}\Xi^\prime_c$), $\bar{D}\Sigma_c^*$ ($\bar{D}\Xi_c^*$), $\bar{D}^{*}\Sigma_c$ ($\bar{D}^{*}\Xi^\prime_c$)  and   $\bar{D}^{*}\Sigma_c^*$ ($\bar{D}^{*}\Xi_c^*$)  pentaquark molecular states with the $J^P={\frac{1}{2}}^-$, ${\frac{3}{2}}^-$, ${\frac{3}{2}}^-$ and ${\frac{5}{2}}^-$, respectively, which lie near  the corresponding  $\bar{D}\Sigma_c$ ($\bar{D}\Xi^\prime_c$), $\bar{D}\Sigma_c^*$ ($\bar{D}\Xi_c^*$), $\bar{D}^{*}\Sigma_c$ ($\bar{D}^{*}\Xi^\prime_c$)  and   $\bar{D}^{*}\Sigma_c^*$ ($\bar{D}^{*}\Xi_c^*$) thresholds, respectively, see Table \ref{mass-pole-residue-tab}.
 The two-body strong decays to the corresponding open-charm meson-baryon pairs, such as $\bar{D}\Sigma_c$ ($\bar{D}\Xi^\prime_c$), $\bar{D}\Sigma_c^*$ ($\bar{D}\Xi_c^*$), $\bar{D}^{*}\Sigma_c$ ($\bar{D}^{*}\Xi^\prime_c$)  and   $\bar{D}^{*}\Sigma_c^*$ ($\bar{D}^{*}\Xi_c^*$), with the fall-apart mechanism directly,  can only take place through the higher tails of the mass distributions and are kinematically suppressed in the phase space, the widths of those pentaquark molecular states should be narrow. The large width
 $\Gamma_{P_c(4380)}=205\pm 18\pm 86\,\rm{MeV}$ maybe indicate that the $P_c(4390)$ maybe correspond to two or more unresolved structures. More experimental data and theoretical works  are still needed to identify  the $P_c(4312)$, $P_c(4380)$, $P_c(4440)$, $P_c(4457)$ and $P_{cs}(4459)$ unambiguously.

 In the present work, we make predictions for the masses of new pentaquark molecular states besides reproducing the masses of the existing pentaquark candidates $P_c(4312)$, $P_c(4380)$, $P_c(4440)$, $P_c(4457)$ and $P_{cs}(4459)$.  We can search for the non-strange $\bar{D}\Sigma_c$, $\bar{D}\Sigma_c^*$, $\bar{D}^{*}\Sigma_c$  and  $\bar{D}^{*}\Sigma_c^*$  pentaquark molecular states  with the isospin $I=\frac{1}{2}$ (or $I=\frac{3}{2}$) and with the spin-parity $J^P={\frac{1}{2}}^-$, ${\frac{3}{2}}^-$, ${\frac{3}{2}}^-$ and ${\frac{5}{2}}^-$, respectively  in the $\Lambda_b^0$ decays,
 \begin{eqnarray}
\Lambda_b^0 &\to&   p     J/\psi K^-\, , \, n     J/\psi \bar{K}^0\, , \,n    J/\psi \bar{K}^0\, , \,p     \eta_c K^-\, , \, n     \eta_c \bar{K}^0\, , \,n    \eta_c \bar{K}^0\, , \nonumber\\
&&  \Delta^+     J/\psi K^-\, , \, \Delta^0     J/\psi \bar{K}^0\, , \,\Delta^0    J/\psi \bar{K}^0\, , \,\Delta^+     \eta_c K^-\, , \, \Delta^0     \eta_c \bar{K}^0\, , \,\Delta^0   \eta_c \bar{K}^0\, ,
\end{eqnarray}
and search for the strange $\bar{D}\Xi^\prime_c$, $\bar{D}\Xi_c^*$, $\bar{D}^{*}\Xi^\prime_c$  and   $\bar{D}^{*}\Xi_c^*$  pentaquark molecular states with the isospin $I=0$ (or $I=1$) with the spin-parity $J^P={\frac{1}{2}}^-$, ${\frac{3}{2}}^-$, ${\frac{3}{2}}^-$ and ${\frac{5}{2}}^-$, respectively  in the $\Xi_b^0$ and $\Xi_b^-$ decays,
 \begin{eqnarray}
\Xi_b^0 &\to&   \Sigma^+     J/\psi K^-\, , \, \Sigma^0     J/\psi \bar{K}^0\, , \,\Lambda^0     J/\psi \bar{K}^0\, , \, \Sigma^+     \eta_c K^-\, , \, \Sigma^0     \eta_c \bar{K}^0\, , \,\Lambda^0    \eta_c \bar{K}^0\, ,\nonumber\\
&&\Sigma^{*+}     J/\psi K^-\, , \, \Sigma^{*0}     J/\psi \bar{K}^0\, , \, \Sigma^{*+}     \eta_c K^-\, , \, \Sigma^{*0}     \eta_c \bar{K}^0\, ,
\end{eqnarray}
\begin{eqnarray}
\Xi_b^- &\to&  \Lambda^0     J/\psi K^-\, , \, \Sigma^0     J/\psi K^-\, , \, \Sigma^-     J/\psi \bar{K}^0\, , \,\Lambda^0     \eta_c K^-\, , \, \Sigma^0     \eta_c K^-\, , \, \Sigma^-     \eta_c \bar{K}^0\, , \nonumber\\
&&\Sigma^{*0}     J/\psi K^-\, , \, \Sigma^{*-}     J/\psi \bar{K}^0\, , \, \Sigma^{*0}     \eta_c K^-\, , \, \Sigma^{*-}     \eta_c \bar{K}^0\, .
\end{eqnarray}

\begin{table}
\begin{center}
\begin{tabular}{|c|c|c|c|c|c|c|c|}\hline\hline
                                 &$J^P$                 &$\mu(\rm GeV)$    &$T^2 (\rm{GeV}^2)$  &$\sqrt{s_0}(\rm{GeV})$   &pole     \\ \hline

$\bar{D}\,\Sigma_c$              &${\frac{1}{2}}^-$     &2.2               &$3.1-3.5$           &$5.02\pm0.10$            &$(42-64)\%$  \\   \hline

$\bar{D}\,\Xi_c^{\prime}$        &${\frac{1}{2}}^-$     &2.2               &$3.2-3.6$           &$5.14\pm0.10$            &$(42-63)\%$  \\   \hline

$\bar{D}\,\Sigma_c^{*} $         &${\frac{3}{2}}^-$     &2.4               &$3.2-3.6$           &$5.08\pm0.10$            &$(43-64)\%$  \\    \hline

$\bar{D}\,\Xi_c^{*} $            &${\frac{3}{2}}^-$     &2.4               &$3.3-3.7$           &$5.21\pm0.10$            &$(43-64)\%$  \\    \hline

$\bar{D}^{*}\,\Sigma_c $         &${\frac{3}{2}}^-$     &2.5               &$3.3-3.7$           &$5.16\pm0.10$            &$(41-62)\%$  \\   \hline

$\bar{D}^{*}\,\Xi_c^{\prime}$    &${\frac{3}{2}}^-$     &2.5               &$3.4-3.8$           &$5.29\pm0.10$            &$(41-61)\%$  \\   \hline

$\bar{D}^{*}\,\Sigma_c^{*}$      &${\frac{5}{2}}^-$     &2.6               &$3.4-3.8$           &$5.22\pm0.10$            &$(40-60)\%$  \\ \hline

$\bar{D}^{*}\,\Xi_c^{*}$         &${\frac{5}{2}}^-$     &2.6               &$3.5-3.9$           &$5.35\pm0.10$            &$(40-60)\%$  \\ \hline
\hline
\end{tabular}
\end{center}
\caption{ The  optimal energy scales $\mu$, Borel parameters $T^2$, continuum threshold parameters $s_0$ and
 pole contributions (pole)    for the hidden-charm   pentaquark molecular states.} \label{Borel-pole}
\end{table}

\begin{table}
\begin{center}
\begin{tabular}{|c|c|c|c|c|c|c|c|}\hline\hline
                              &$J^P$                 &$M (\rm{GeV})$     &$\lambda (10^{-3}\rm{GeV}^6)$ & Thresholds (MeV) & Assignments        \\  \hline

$\bar{D}\,\Sigma_c$           &${\frac{1}{2}}^-$     &$4.32\pm0.12$      &$2.00\pm0.36   $              &4318              & ? $P_c(4312)$     \\      \hline

$\bar{D}\,\Xi_c^{\prime}$     &${\frac{1}{2}}^-$     &$4.45\pm0.12$      &$2.32\pm0.42   $              &4443              & ? $P_{cs}(4459)$  \\      \hline

$\bar{D}\,\Sigma_c^{*} $      &${\frac{3}{2}}^-$     &$4.38\pm0.12$      &$1.25\pm0.21  $               &4382              &  ? $P_c(4380)$  \\  \hline

$\bar{D}\,\Xi_c^{*} $         &${\frac{3}{2}}^-$     &$4.51\pm0.11$      &$1.45\pm0.25  $               &4510              &  ? ? $P_{cs}(4459)$     \\  \hline

$\bar{D}^{*}\,\Sigma_c$       &${\frac{3}{2}}^-$     &$4.46\pm0.12$      &$2.37\pm0.40 $                &4460              &  ? $P_c(4440/4457)$    \\ \hline

$\bar{D}^{*}\,\Xi_c^{\prime}$ &${\frac{3}{2}}^-$     &$4.60\pm0.11$      &$2.80\pm0.48 $                &4585              &     \\ \hline

$\bar{D}^{*}\,\Sigma_c^{*}$   &${\frac{5}{2}}^-$     &$4.52\pm 0.12$     &$1.82\pm0.31 $                &4524              &  ? ? $P_c(4457)$    \\ \hline

$\bar{D}^{*}\,\Xi_c^{*}$      &${\frac{5}{2}}^-$     &$4.67\pm 0.11$     &$2.16\pm0.37 $                &4652              &          \\ \hline
\hline
\end{tabular}
\end{center}
\caption{ The predicted masses and pole residues of the hidden-charm   pentaquark molecular states with the possible assignments, where the double-? denotes that such assignments are not excluded due to the uncertainties.} \label{mass-pole-residue-tab}
\end{table}

\section{Conclusion}
In this article, we investigate the $\bar{D}\Sigma_c$, $\bar{D}\Xi^\prime_c$,    $\bar{D}\Sigma_c^*$, $\bar{D}\Xi_c^*$, $\bar{D}^{*}\Sigma_c$, $\bar{D}^{*}\Xi^\prime_c$,
  $\bar{D}^{*}\Sigma_c^*$ and $\bar{D}^{*}\Xi_c^*$ pentaquark molecular states with strangeness and without strangeness via  the QCD sum rules at length by carrying out the operator product expansion   up to   the vacuum condensates of dimension $13$ in a consistent way, and take the modified energy scale formula to choose the best energy scales of the spectral densities at the quark-gluon level, and make predictions for the masses of new pentaquark molecular states besides reproducing the masses of the existing pentaquark candidates $P_c(4312)$, $P_c(4380)$, $P_c(4440)$, $P_c(4457)$ and $P_{cs}(4459)$.    The present calculations  support assigning the $P_c(4312)$ as the $\bar{D}\Sigma_c$ pentaquark molecular state with the quantum numbers  $J^P={\frac{1}{2}}^-$ and $I=\frac{1}{2}$, assigning the $P_c(4380)$ as the $\bar{D}\Sigma_c^*$ pentaquark molecular state with the quantum numbers $J^P={\frac{3}{2}}^-$  and $I=\frac{1}{2}$,  assigning the $P_c(4440/4457)$ as the $\bar{D}^{*}\Sigma_c$ pentaquark molecular state with the quantum numbers $J^P={\frac{3}{2}}^-$  and $I=\frac{1}{2}$,  assigning the $P_{cs}(4459)$ as the $\bar{D}\Xi^\prime_c$ pentaquark molecular state with the quantum numbers $J^P={\frac{1}{2}}^-$  and $I=0$; but cannot
  exclude the possibilities of assigning the $P_c(4457)$ as the $\bar{D}^{*}\Sigma_c^*$ pentaquark molecular state with the quantum numbers $J^P={\frac{5}{2}}^-$  and $I=\frac{1}{2}$ and   assigning the $P_{cs}(4459)$ as the $\bar{D}\Xi^*_c$ pentaquark molecular state with the quantum numbers $J^P={\frac{3}{2}}^-$  and $I=0$ due to the uncertainties.
   In calculations, we observe that  the predicted masses of the pentaquark molecular states without strangeness and with strangeness have mass gap about
   $ 0.13\sim0.15\,\rm{GeV}$, which is consistent with the light-flavor $SU(3)$ breaking effects of the heavy baryons in the flavor sextet ${\bf 6}_f$.
     We can search for both the old and new pentaquark molecular states in the decays of the $\Lambda_b^0$, $\Xi_b^0$ and $\Xi_b^-$ in the future to preform more robust investigations and shed light on the nature of the $P_c$ and $P_{cs}$ states.

\section*{Acknowledgements}
This  work is supported by National Natural Science Foundation, Grant Number  11775079.


\begin{thebibliography}{99}

\bibitem{LHCb-4380} R. Aaij  et al, Phys. Rev. Lett. {\bf 115} (2015) 072001.

\bibitem{LHCb-Pc4312} R. Aaij et al, Phys. Rev. Lett. {\bf 122} (2019) 222001.


\bibitem{LHCb-Pcs4459-2012} R. Aaij  et al,     Sci. Bull. {\bf 66} (2021) 1278.

\bibitem{PDG}  P. A. Zyla et al,  Prog. Theor. Exp. Phys. {\bf 2020} (2020) 083C01.


\bibitem{Penta-mole-CREFT} M. Z. Liu, Y. W. Pan, F. Z. Peng, M. S. Sanchez, L. S. Geng, A. Hosaka and M. P. Valderrama, Phys. Rev. Lett. {\bf 122} (2019)  242001.

\bibitem{Pcs4459-mole-CREFT} M. Z. Liu, Y. W. Pan and L. S. Geng, Phys. Rev. {\bf D103} (2021)  034003.


\bibitem{Penta-mole-OBE-CC} R. Chen, Z. F. Sun, X. Liu and S. L. Zhu, Phys. Rev. {\bf D100} (2019)  011502.

\bibitem{Pcs4459-mole-OBE-CC} R. Chen, Eur. Phys. J. {\bf C81} (2021)  122.

\bibitem{Pcs4459-mole-OBE-2-CC}  F. L. Wang, X. D. Yang, R. Chen, X. Liu,   Phys. Rev. {\bf D103} (2021)  054025.


\bibitem{Penta-mole-BSE-CC} J. He, Eur. Phys. J. {\bf C79} (2019)  393.

\bibitem{Pcs4459-mole-BSE-HJ-CC} J. T. Zhu, L. Q. Song and J. He,     Phys. Rev. {\bf D103} (2021)  074007.

\bibitem{Pcs4459-mole-BSE-CC} X. K. Dong, F. K. Guo and B. S. Zou, Progr. Phys. {\bf 41} (2021) 65.

\bibitem{Pcs4459-mole-BSE-2-CC} C. W. Xiao, J. J. Wu and B. S. Zou,     Phys. Rev. {\bf D103} (2021)  054016.



\bibitem{Penta-mole-ELA}  C. J. Xiao, Y. Huang, Y. B. Dong, L. S. Geng and D. Y. Chen, Phys. Rev. {\bf D100} (2019)  014022.

\bibitem{Pcs4459-mole-ELA} Q. Wu, D. Y. Chen and R. Ji,    Chin. Phys. Lett. {\bf 38} (2021) 071301.

\bibitem{Penta-mole-ERE-CC} Z. H. Guo and J. A. Oller, Phys. Lett. {\bf B793} (2019) 144.

\bibitem{Penta-mole-LSE-CC} F. K. Guo, H. J. Jing, U. G. Meissner and S. Sakai, Phys. Rev. {\bf D99} (2019) 091501.

\bibitem{Penta-mole-LSE-2-CC} M. L. Du, V. Baru, F. K. Guo, C. Hanhart, U. G. Meissner, J. A. Oller and Q. Wang, Phys. Rev. Lett. {\bf 124} (2020)  072001.


\bibitem{Penta-mole-QCDSR-Chen} H. X. Chen, W. Chen and S. L. Zhu, Phys. Rev. {\bf D100} (2019) 051501.

\bibitem{Penta-mole-QCDSR-Zhang} J. R. Zhang, Eur. Phys. J. {\bf C79} (2019)  1001.

\bibitem{Penta-mole-QCDSR-Azizi} K. Azizi, Y. Sarac and H. Sundu, Phys. Rev. {\bf D95} (2017)  094016.





\bibitem{WZG-IJMPA-mole} Z. G. Wang, Int. J. Mod. Phys. {\bf A34} (2019)  1950097.


\bibitem{Pcs4459-mole-QCDSR} H. X. Chen, W. Chen and X. Liu and X. H. Liu,     Eur. Phys. J. {\bf C81} (2021)  409.


\bibitem{WZG-color-neutral} Z. G. Wang, arXiv:2103.04236 [hep-ph].


\bibitem{di-di-anti-penta-mass-1} R. Ghosh, A. Bhattacharya and B. Chakrabarti, Phys. Part. Nucl. Lett. {\bf 14} (2017)  550.

\bibitem{di-di-anti-penta-mass-2} V. V. Anisovich, M. A. Matveev, J. Nyiri, A. V. Sarantsev and A. N. Semenova, Int. J. Mod. Phys. {\bf A30} (2015) 1550190.


\bibitem{di-tri-penta-mass} R. Zhu and C. F. Qiao, Phys. Lett. {\bf B756} (2016) 259.


\bibitem{di-di-anti-penta-decay-1} L. Maiani, A. D. Polosa and V. Riquer,  Phys. Lett. {\bf B749} (2015) 289.

\bibitem{di-di-anti-penta-decay-2} G. N. Li, M. He and X. G. He,  JHEP {\bf 1512} (2015) 128.

\bibitem{di-di-anti-penta-decay-3} H. Y. Cheng and C. K. Chua, Phys. Rev. {\bf D92} (2015) 096009.

\bibitem{di-tri-penta-width} R. F. Lebed, Phys. Lett. {\bf B749} (2015) 454.


\bibitem{Wang1508-EPJC} Z. G. Wang, Eur. Phys. J. {\bf C76} (2016) 70.

\bibitem{WangHuang1508-1} Z. G. Wang  and T. Huang, Eur. Phys. J. {\bf C76} (2016)  43.

\bibitem{WangHuang1508-2} Z. G. Wang, Eur. Phys. J. {\bf C76} (2016)  142.

\bibitem{WangHuang1508-3} Z. G. Wang, Nucl. Phys. {\bf B913} (2016) 163.


\bibitem{WZG-IJMPA-penta} Z. G. Wang, Int. J. Mod. Phys. {\bf A35} (2020) 2050003.

\bibitem{WZG-Pcs4459}  Z. G. Wang, Int. J. Mod. Phys. {\bf A36} (2021)  2150071.


\bibitem{Pcs4459-penta-di-di-decay} K. Azizi, Y. Sarac and H. Sundu,     Phys. Rev. {\bf D103} (2021)  094033.

\bibitem{Pcs4459-penta-di-di-EM} U. Ozdem, Eur. Phys. J. {\bf C81} (2021)  277.


\bibitem{WZG-WHJ-XQ} Z. G. Wang, H. J. Wang and Q. Xin, Chin. Phys. {\bf C45} (2021) 063104.


\bibitem{Can-Sigmac-2015} K. U. Can, G. Erkol, B. Isildak, M. Oka and T. Takahashi, PoS LATTICE2014 (2015) 157.


\bibitem{Kim-Radii-soliton} J. Y. Kim and H. C. Kim, Phys. Rev. {\bf D97} (2018) 114009.

\bibitem{Hwang-Radii-LF} C. W. Hwang, Eur. Phys. J. {\bf C23} (2002) 585.



\bibitem{Wang-tetraquark-QCDSR-1} Z. G. Wang and T. Huang, Phys. Rev. {\bf D89} (2014) 054019.

\bibitem{Wang-tetraquark-QCDSR-2} Z. G. Wang, Eur. Phys. J. {\bf C74} (2014)  2874.

\bibitem{Wang-tetraquark-QCDSR-3} Z. G. Wang and T. Huang, Nucl. Phys. {\bf A930} (2014) 63.

\bibitem{Wang-tetraquark-QCDSR-4} Z. G. Wang and Y. F. Tian, Int. J. Mod. Phys. {\bf A30} (2015) 1550004.



\bibitem{Wang-molecule-QCDSR-1} Z. G. Wang and T. Huang, Eur. Phys. J. {\bf C74} (2014)  2891.

\bibitem{Wang-molecule-QCDSR-2} Z. G. Wang, Eur. Phys. J. {\bf C74} (2014)  2963.



\bibitem{GMOR} M. Gell-Mann, R. J. Oakes and B. Renner, Phys. Rev. {\bf 175} (1968) 2195.



\bibitem{WZG-DvDvDv} Z. G. Wang, Commun. Theor. Phys. {\bf 73} (2021)  065201.



\bibitem{Review-rho-varrho} D.  B. Leinweber, Annals Phys. {\bf 254} (1997) 328; and references  therein.

\bibitem{Narison-rho} S. Narison,  Phys. Lett. {\bf B673} (2009) 30.



\bibitem{SVZ79}  M. A. Shifman, A. I. Vainshtein and V. I. Zakharov, Nucl. Phys. {\bf B147} (1979) 385, 448.

\bibitem{PRT85} L. J. Reinders, H. Rubinstein and S. Yazaki, Phys. Rept. {\bf 127} (1985) 1.

\bibitem{ColangeloReview} P. Colangelo and A. Khodjamirian, hep-ph/0010175.



\bibitem{Narison-mix} S. Narison and R. Tarrach, Phys. Lett. {\bf 125 B} (1983) 217.


\bibitem{Wang-CPC-4390} Z. G. Wang, Chin. Phys. {\bf C41} (2017)  083103.


\bibitem{Cornell} E. J. Eichten and C. Quigg, Phys. Rev. {\bf D99} (2019) 054025.


\bibitem{WZG-hidden-charm-mole} Z. G. Wang,   Int. J. Mod. Phys. {\bf A35} (2021) 2150107.


\end{thebibliography}
\end{document}